\documentclass[]{aa}
\usepackage[varg]{txfonts}
\usepackage{natbib}
\usepackage[final]{hyperref}
\usepackage{xcolor}
\usepackage{multirow}
\usepackage[switch]{lineno}
\usepackage{booktabs}
\usepackage{siunitx}
\usepackage{upgreek}
\usepackage[normalem]{ulem} 

\begin{document}

\title{Galaxy and Mass Assembly (GAMA)}
\subtitle{Mid-infrared properties as tracers of galaxy environment}

\author{
        U.~Sureshkumar \inst{\ref{aff:oauj}} 
        \and A.~Durkalec \inst{\ref{aff:ncbj}} 
        \and A.~Pollo \inst{\ref{aff:oauj},\ref{aff:ncbj}}
        \and M.~Bilicki \inst{\ref{aff:ctp_warsaw}}
        \and M.~E.~Cluver \inst{\ref{aff:cas_swisburne}, \ref{aff:phynastro_westcape}}
        \and S.~Bellstedt \inst{\ref{aff:icrar_australia}}
        \and D.~J.~Farrow \inst{\ref{mpe_garching}, \ref{lmu_munich}}
        \and J.~Loveday \inst{\ref{aff:sussex}}
        \and E.~N.~Taylor \inst{\ref{aff:cas_swisburne}}
        \and J.~Bland-Hawthorn \inst{\ref{aff:sydneyInstForAstronomy_Australia}, \ref{aff:centreOfExcellence_Australia}}
        }

\institute{
    Astronomical Observatory of the Jagiellonian University, ul. Orla 171, 30-244 Krak\'{o}w, Poland \\\email{usureshkumar@oa.uj.edu.pl} 
    \label{aff:oauj}
    \and
    National Centre for Nuclear Research, ul. Pasteura 7, 02-093 Warsaw, Poland 
    \label{aff:ncbj}
    \and
    Center for Theoretical Physics, Polish Academy of Sciences, al. Lotnik\'{o}w 32/46, 02-668 Warsaw, Poland
    \label{aff:ctp_warsaw}
    \and
    Centre for Astrophysics and Supercomputing, Swinburne University of Technology, John Street, Hawthorn, 3122, Australia
    \label{aff:cas_swisburne}
    \and
    Department of Physics and Astronomy, University of the Western Cape, Robert Sobukwe Road, Bellville, 7535, Republic of South Africa
    \label{aff:phynastro_westcape}
    \and
    ICRAR, The University of Western Australia, 7 Fairway, Crawley WA 6009, Australia
    \label{aff:icrar_australia}
    \and
    Max Planck Institute for Extraterrestrial Physics, Gie{\ss}enbachstra{\ss}e, 85748, Garching b. M\"{u}nchen, Germany
    \label{mpe_garching}
    \and
    Universit{\"a}ts-Sternwarte, Fakult{\"a}t f{\"u}r Physik, Ludwig-Maximilians-Universit{\"a}t M{\"u}nchen, Scheinerstr. 1, 81679 M{\"u}nchen, Germany
    \label{lmu_munich}
    \and
    Astronomy Centre, University of Sussex, Falmer, Brighton BN1 9QH, UK
    \label{aff:sussex}
    \and
    Sydney Institute for Astronomy, School of Physics, University of Sydney, NSW 2006, Australia
    \label{aff:sydneyInstForAstronomy_Australia}
    \and
    Centre of Excellence for All-Sky Astrophysics in 3D, Australia
    \label{aff:centreOfExcellence_Australia}
    }

\date{Received 25 January 2022 / Accepted 02 November 2022}

\abstract
{}
{We investigate how different mid-infrared (mid-IR) properties of galaxies are correlated with the environment in which the galaxies are located. 
For this purpose, we first study the dependence of galaxy clustering on the absolute magnitude at 3.4 $\mu\mathrm{m}$ and redshift.
Then, we look into the environmental dependence of mid-IR luminosities and the galaxy properties derived from these luminosities.
We also explore how various IR galaxy luminosity selections influence the galaxy clustering measurements.
}
{We used a set of W1 (3.4 $\mu\mathrm{m}$) absolute magnitude ($M_\text{W1}$) selected samples from the Galaxy and Mass Assembly (GAMA) survey matched with mid-IR properties from the Wide-field Infrared Survey Explorer (WISE) in the redshift range $0.07 \leq z < 0.43$.
We computed the galaxy two-point correlation function (2pCF) and compared the clustering lengths between subsamples binned in $M_\text{W1}$ and in redshift. 
We also measured the marked correlation function (MCF), in which the galaxies are weighted by marks when measuring clustering statistics, using the luminosities in the WISE W1 to W4 (3.4 to 22 $\mu\mathrm{m}$) bands as marks.
Additionally, we compared the measurements of MCFs with different estimates of stellar mass and star formation rate (SFR) used as marks. 
Finally, we checked how different selections applied to the sample affect the clustering measurements.
}
{We show strong clustering dependence on the W1 absolute magnitude: galaxies brighter in the W1 band are more strongly clustered than their fainter counterparts. 
We also observe a lack of significant redshift dependence of clustering in the redshift range $0.07 \leq z < 0.43$.
We show that although the W1 and W2 bands are direct indicators of stellar mass, a galaxy sample selected based on W1 or W2 bands does not perfectly show the clustering behaviour of a stellar mass-selected sample.  
The proxy relation between W3 and W4 bands and SFR is similar.
We also demonstrate the influence of estimation techniques of stellar mass and SFR on the clustering measurements.
}
{}


\titlerunning{Mid-infrared properties as tracers of galaxy environment in GAMA}
\authorrunning{U. Sureshkumar et al.}

\maketitle

\section{Introduction}\label{sec:introduction}

According to the $\Lambda$ cold dark matter ($\Lambda$CDM) model, the large-scale structure (LSS) of the Universe resulted from the evolution of primordial fluctuations in the matter density distribution under the influence of gravity \citep{springel2005}.
During the evolution, the dark matter started clumping to form dark matter haloes that provided the potential wells for the formation of galaxies \citep{press_schechter_1974, white&rees1978}.
As a result, the galaxies live in dark matter haloes, and the halo properties are expected to have a significant influence on the galaxy properties. 
Hence, the environmental dependence of halo properties prompts a correlation between galaxy properties and the environment in which they live \citep{wechsler2018}.
Therefore, deeper insights into the environmental dependence of galaxy properties and their redshift evolution are crucial for understanding connections between dark matter haloes and their galaxies \citep[see][for a review]{somerville2015} and might provide useful constraints on our understanding of galaxy formation and evolution.

The dependence of galaxy properties on the environment is imprinted on the observations of galaxy clustering.
One of the methods used to quantify the clustering is the two-point correlation function \citep[2pCF;][]{peebles1980}.
Various studies have used 2pCF to show the dependence of galaxy clustering on different properties such as luminosity \citep{zehavi2011, farrow2015_gama_cf, pollo2006}, stellar mass \citep{skibba2015, durkalec2018}, star formation rate \citep[SFR;][]{hartley2010, lin2012, mostek2013}, colour \citep{coil2008, coupon2012, skibba2014_combining_fields}, and spectral type \citep{norberg2002, meneux2006}.
It is observed that in general, galaxies that are more luminous, massive, redder, and evolved prefer to exist in the denser regions of the LSS than their counterparts. 

Moreover, it is also observed that the amplitude of galaxy clustering depends on the photometric passband in which the volume-limited samples are selected. 
Clustering studies were conducted with galaxy samples selected based on ultraviolet \citep[UV;][]{milliard2007}, optical, such as $B$-band \citep{marulli2013}, $r$-band \citep{zehavi2011}, and $g$-band \citep{skibba2014_combining_fields}, and infrared \citep[IR;][]{sobral2010, oliver2004, solarz2015, pollo2013_akarifull} observations.
These studies concluded that galaxies that are luminous in the optical and IR bands exhibit stronger clustering and reside in the denser environment of the LSS than their fainter counterparts. 
However, galaxies that are luminous in the $u$ band tend to reside in less dense environments of the LSS \citep{deng2012}.
All these studies revealed that different galaxy properties correlate differently with the clustering, and hence the environment. 

A more efficient tool than the 2pCF for detecting the environmental dependence of galaxy properties is the marked correlation function \citep[MCF;][]{sheth&tormen2004, skibba2013}.
The marked correlation function, defined for a given property (referred to as \textit{\textup{mark}}) is computed by weighting the galaxies by that property during the clustering measurement (see Sect.~\ref{sec:measurement_mcf} for details). 
Quite a number of studies have used MCF to study the environmental dependence of galaxy properties such as luminosity, stellar mass, colour, age, morphology, SFR, and spin
: \citet{beisbart2000}, \citet{skibba2006}, \citet{skibba2013}, \citet{sheth2006}, \citet{gunawardhana2018}, \citet{sureshkumar2021}, and \citet{rutherford2021}.
Additionally, \citet{riggs2021} used the MCF with group mass as a mark to study the clustering of galaxy groups.
The MCF is also used to constrain modified gravity theories \citep{white2016_mcf_gravity, armijo2018_mcf_gravity, hernandez-aguayo2018, satpathy2019_mcf_gravity, alam2020_mcf_desi}.
Recently, \citet{sureshkumar2021} used MCFs to study how properties such as luminosities in $u, g, r, J, K$ bands, stellar mass, SFR, and specific SFR trace small-scale clustering in the Galaxy and Mass Assembly \citep[GAMA;][]{driver2009_gama_gen} survey.
We observed a hierarchy of luminosity passbands in which the redder $K$ band appears to trace the galaxy clustering better than $u, g, r,$ and $J$ bands.

In this context, it is interesting to explore how even longer wavelengths (IR) are correlated with the galaxy clustering. 
The clustering of IR galaxies is of particular interest because different parts of the IR spectrum trace different physical processes.
The near-IR emission of a galaxy primarily arises from the evolved stellar populations and hence serves as a good indicator of its stellar mass  \citep{kauffmann1998, cole2001}.
The mid-IR region has a dual capability: lower wavelengths are sensitive to light from the evolved population of stars, and higher wavelengths are sensitive to star formation \citep{meidt2012, wen2013, jarrett2013, cluver2014}.
The far-IR region has also been proven to be a good indicator of SFR \citep{calzetti2010}.

There have been studies of clustering of galaxies selected in near-IR wavelengths \citep[e.g.][]{baugh1996, roche1998, roche1999, daddi2000, kummel2000, maller2005}.
The clustering of mid-IR selected galaxies is also well studied using surveys and telescopes such as the Infrared Astronomical Satellite \citep[IRAS,][]{fisher1994}, the European Large-Area \textit{ISO} survey \citep[ELAIS;][]{gonzalez-solares2004, delia2005}, and \textit{Spitzer} \citep{fang2004, oliver2004, gilli2007, torre2007, waddington2007}.
In the far-IR region, clustering studies are conducted with Herschel \citep{cooray2010, maddox2010, magliocchetti2011} and AKARI \citep{pollo2013_akarinorth, pollo2013_akarifull}.
It is observed that in general, the IR-selected galaxies show stronger clustering than optically selected ones.

In the mid-IR regime, the Wide-field Infrared Survey Explorer \citep[WISE;][]{wright2010_wise} covers the entire sky at four bands:   W1 (3.4~$\mu\mathrm{m}$), W2 (4.6~$\mu\mathrm{m}$), W3 (12~$\mu\mathrm{m}$), and W4 (22~$\mu\mathrm{m}$).
The W1 and W2 bands, being closer to near-IR bands, trace the continuum emission from evolved stars with minimum extinction.
The W3 and W4 bands are good indicators of the interstellar medium and star formation activity: W3 is dominated by the stochastically heated 11.3~$\mu\mathrm{m}$ polycyclic aromatic hydrocarbon (PAH) and 12.8~$\mu\mathrm{m}$ [Ne II] emission features, and W4 traces the dust continuum that is a combination of warm and cold small grains in equilibrium \citep{jarrett2013, cluver2014}.
As different WISE bands trace different galaxy properties, WISE photometric data can be useful to study how various mid-IR properties follow the galaxy environment.

The main aim of this paper is to use the 2pCFs and MCFs to explore the environmental dependence of WISE band properties in GAMA. 
We use a set of galaxy samples from the GAMA survey enhanced with mid-IR properties from WISE \citep{cluver2014, cluver2020}. 
Because of its high completeness (> 98.5\%) down to $r < 19.8$, GAMA has been abundantly used for clustering studies \citep[e.g.][]{christodoulou2012_gama_angular, lindsay2014, farrow2015_gama_cf, loveday2018_gama_pvd, gunawardhana2018, sureshkumar2021}.
In particular, using the WISE photometry, the clustering of galaxies in the G12 region of GAMA was studied by \citet{jarrett2017}.
In our work, we study the correlations of the WISE luminosities with galaxy clustering in all three equatorial regions of GAMA (G09, G12, and G15) using MCFs.
We also explore the dependence of galaxy clustering on the W1 absolute magnitude and redshift using 2pCFs. 

Additionally, we check how different methods of estimating galaxy properties may influence the measurements of MCFs.
To do this, we use three independent estimates of stellar mass in GAMA: one from stellar population synthesis (SPS) modelling by \citet{taylor2011_gama_stellar_mass}, a second estimate based on mid-IR photometry by \citet{cluver2014}, and a third derived using the \textsc{ProSpect} code \citep{robotham2020}.
Similarly, we repeat the comparison using independent SFR estimates.
In addition to the \textsc{MagPhys} SFR \citep{dacunha2008_gama_sfr}, SFRs based on the W3 and W4 bands by \citet{cluver2017} and from \textsc{ProSpect} code are available.
The \citet{taylor2011_gama_stellar_mass} and \citet{cluver2014} stellar masses were checked for consistency by \citet{kettlety2018}.
In this work, we measure and compare MCFs using these different estimates as marks to determine how well they agree.

The paper is organised as follows. 
In Sect.~\ref{sec:data} we describe properties of the GAMA survey, GAMA-WISE, and ProSpect catalogues, our sample selection methods, and the random catalogue. 
Sect.~\ref{sec:measurement} describes different clustering techniques and their definitions. 
The results of our measurements are presented in Sect.~\ref{sec:results}, are discussed and compared with other works in Sect.~\ref{sec:discussion}, and we finally conclude in Sect.~\ref{sec:conclusions}. 

Throughout the paper, a flat $\Lambda$CDM cosmological model with $\Omega_{\text{M}}=0.3$ and $\Omega_\Lambda=0.7$ is adopted, and the Hubble constant is parametrised via $h=H_0/100 \rm \, km \, s^{-1} \, Mpc^{-1}$. 
All galaxy properties except for those from the ProSpect catalogue are measured using $h=0.7$.
The ProSpect catalogue was created using $h=0.678$ \citep{bellstedt2020_sed}.
The distances are expressed in comoving coordinates and are in units of $h^{-1} \mathrm{Mpc}$.

\section{Data}\label{sec:data}

\subsection{Galaxy and Mass Assembly}\label{sec:data_gama}

Galaxy and Mass Assembly is a spectroscopic survey that observes galaxies down to extinction-corrected $r$-band Petrosian magnitudes $r_\text{petro} < 19.8$ mag.
The multi-wavelength coverage of GAMA provides a sampling of the UV to far-IR range of wavelengths (0.15~--~500~$\mu\mathrm{m}$) through 21 broad-band filters: far-UV and near-UV \citep[\textit{GALEX};][]{martin2005_galex}, \textit{ugri} \citep[KiDS;][]{dejong2013_kids}, $ZYJHK_s$ \citep[VIKING;][]{edge2013_viking}, W1, W2, W3, and W4 \citep[WISE;][]{wright2010_wise}, and 100~$\mu\mathrm{m}$, 160~$\mu\mathrm{m}$, 250~$\mu\mathrm{m}$, 350~$\mu\mathrm{m}$, and 500~$\mu\mathrm{m}$ \citep[\textit{Herschel-}ATLAS;][]{eales2010_herschel-atlas}. 
The GAMA survey covers three equatorial regions called G09, G12, and G15, and two southern regions G02 and G23, each with 12 $\times$ 5 $\text{deg}^2$ of sky coverage.
Detailed descriptions of GAMA can be found in \citet{driver2009_gama_gen}, \citet{robotham2010_gama_tiling}, \citet{driver2011_gama_coredata}, and \citet{liske2015_gama_dr2}.

As the primary data, we used the $r$-band-limited data from GAMA II equatorial regions (G09, G12, and G15) where the survey provides high spatial completeness and an overall redshift completeness of $98.48\%$. 
We selected GAMA main survey galaxies (\texttt{SURVEY\_CLASS} $\geq$ 4) with spectroscopic redshifts in the range $0.07 < z < 0.43$. 
We chose $z_\text{min} = 0.07$ to avoid our samples being dominated by the local structures, whereas we chose $z_\text{max} = 0.43$ to optimally select populous volume-limited samples.
We only used secure redshifts with quality flag \texttt{nQ} $\geq$  3, which ensures that the redshift is correct to $>90\%$ .
In addition to this redshift quality cut, we only selected objects with \texttt{VIS\_CLASS} = 0, \texttt{VIS\_CLASS} = 1, or \texttt{VIS\_CLASS} = 255 to avoid sources that are visually classified to be deblends of stars or parts of other galaxies \citep{baldry2010}.

\subsection{Galaxy properties}\label{data_properties}

In this work, we first use 2pCF to study the dependence of galaxy clustering on the W1 absolute magnitude and redshift.
Then we use MCFs to explore the environmental dependence of 11 galaxy properties: four luminosities ({\small $L_\text{W1}$}, {\small $L_\text{W2}$}, {\small $L_\text{W3}$}, and {\small $L_\text{W4}$}), three stellar mass estimates ({\small $M^{\star}_\text{GAMA}$}, {\small $M^{\star}_\text{WISE}$}, and {\small $M^{\star}_\text{ProSpect}$}), and four SFR estimates ({\small $\text{SFR}_\text{GAMA}$}, {\small $\text{SFR12}_\text{WISE}$}, {\small $\text{SFR22}_\text{WISE}$}, and {\small $\text{SFR}_\text{ProSpect}$}). 

\subsubsection{GAMA catalogue}\label{data_gamacat}

As a base for comparison, we used the stellar mass ({\small $M^{\star}_\text{GAMA}$}) and star formation rate ({\small $\text{SFR}_\text{GAMA}$}) from the main GAMA catalogue.
{\small $M^{\star}_\text{GAMA}$} is from \textsc{StellarMassesLambdarv20} Data Management Unit (DMU) \citep{taylor2011_gama_stellar_mass,wright2016}.
These stellar mass estimates are based on the methods of \citet{taylor2011_gama_stellar_mass} applied to the \textsc{lambdar} photometry of \citet{wright2016}.
Since the photometry is aperture matched, it is necessary to scale the inferred stellar mass to account for flux lying beyond the
finite aperture. 
To do this, we applied the \texttt{fluxscale} (ratio of the $r$ -band aperture flux and the total S\'{e}rsic flux) correction to {\small $M^{\star}_\text{GAMA}$} as described in \citet{taylor2011_gama_stellar_mass}.
In addition to the filtering described in Sect.~\ref{sec:data_gama}, we retained only those galaxies with $0 < \texttt{fluxscale} < 50$. 
We chose this range to avoid extreme \texttt{fluxscale} corrections and at the same time to avoid reducing the sample size significantly.
We used this set of galaxies as our parent sample. 
{\small $\text{SFR}_\text{GAMA}$} was taken from the \textsc{MagPhysv06} DMU and was estimated using the energy balance spectral energy distribution (SED)-fitting code \textsc{magphys} \citep{dacunha2008_gama_sfr}.

\subsubsection{GAMA-WISE catalogue}\label{data_gamawise}

We took the quantities {\small $L_\text{W1}$}, {\small $L_\text{W2}$}, {\small $L_\text{W3}$}, {\small $L_\text{W4}$}, {\small $M^{\star}_\text{WISE}$}, {\small $\text{SFR12}_\text{WISE}$}, and {\small $\text{SFR22}_\text{WISE}$} from the GAMA-WISE catalogue.
This is a source-matched catalogue developed by cross-matching GAMA galaxies with the WISE All-Sky Catalogue using a $3''$ cone search radius with a match rate of well over 95\% \citep{jarrett2017}.
In addition, the resolved galaxies were carefully extracted, and their WISE fluxes were properly derived \citep{cluver2014}.
The galaxy properties such as {\small $M^{\star}_\text{WISE}$} were derived as a function of W1$-$W2 colour \citep{cluver2014}, and {\small $\text{SFR12}_\text{WISE}$} and {\small $\text{SFR22}_\text{WISE}$} were derived using the W3 and W4 luminosities, respectively \citep{cluver2017}. 

It is to be noted that for a WISE-detected source to be included in the All-Sky catalogue, it should have a signal-to-noise ratio (S/N) > 5 in at least one of the four bands.
This means that it is possible to have galaxies in the GAMA-WISE catalogue with a high S/N in one band, but a low S/N in other bands.
The W1 is the most sensitive WISE band and W4 is the least sensitive one.
We refer to \citet{cluver2014,cluver2020} for a detailed description of the construction of this catalogue.

\subsubsection{ProSpect catalogue}\label{data_prospect}

The ProSpect catalogue was made using \textsc{ProSpect} SED-fitting code \citep{robotham2020} applied to the \textsc{ProFound} photometry from GAMA \citep{bellstedt2020_kids}.
As described in \citet{bellstedt2020_sed}, the SEDs were generated using the stellar population template by \citet{bruzual2003} and the dust attenuation law by \citet{charlot&fall2000}.
{\small $M^{\star}_\text{ProSpect}$} and {\small $\text{SFR}_\text{ProSpect}$} used in this work were taken from this ProSpect catalogue.

\subsection{W1 magnitude and redshift-binned samples}\label{sec:data_sampleselection_2pcf}

\begin{figure}
    \centering
    \includegraphics[width=\linewidth]{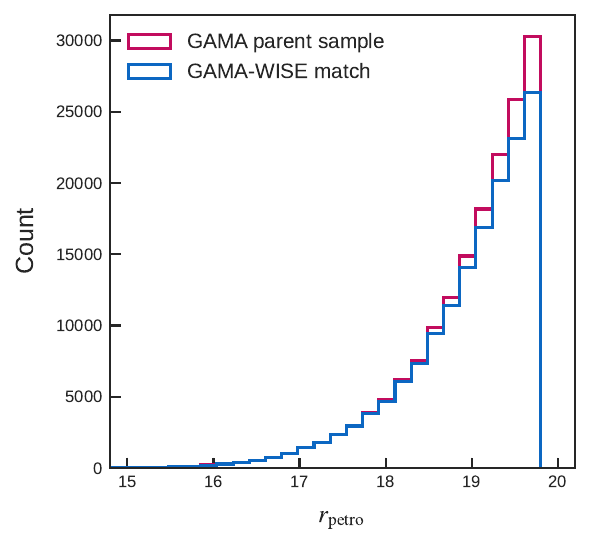}
    \caption{Distribution of the $r$-band flux for the GAMA parent sample and its match sample with the GAMA-WISE catalogue.}
    \label{fig:r_GAMA_WISE}
\end{figure}

As mentioned in Sect.~\ref{data_gamacat}, our parent sample consists of galaxies with a flux limit of $r_\text{petro}~<~19.8$ in the equatorial regions in a redshift range of $0.07 < z < 0.43$.
We selected all galaxies in the parent sample with matches in the GAMA-WISE catalogue and assigned W1 absolute magnitudes by a cross-matching technique using the unique source identifier \texttt{CATAID} of GAMA.
After matching, the sample keeps its completeness. 
About 93\% of the galaxies in the GAMA parent sample have counterparts in the GAMA-WISE catalogue.
In Fig.~\ref{fig:r_GAMA_WISE} we show the distribution of the $r$-band flux for the GAMA and GAMA-WISE samples.
The W1 absolute magnitudes were then corrected for luminosity evolution using the model from \citet{lake2018}, as described in Appendix~\ref{app:lum_evol}.
Then we selected various sets of $r$+W1-selected subsamples using corrected W1 absolute magnitudes. 

\begin{figure*}
    \centering
    \includegraphics[width=\linewidth]{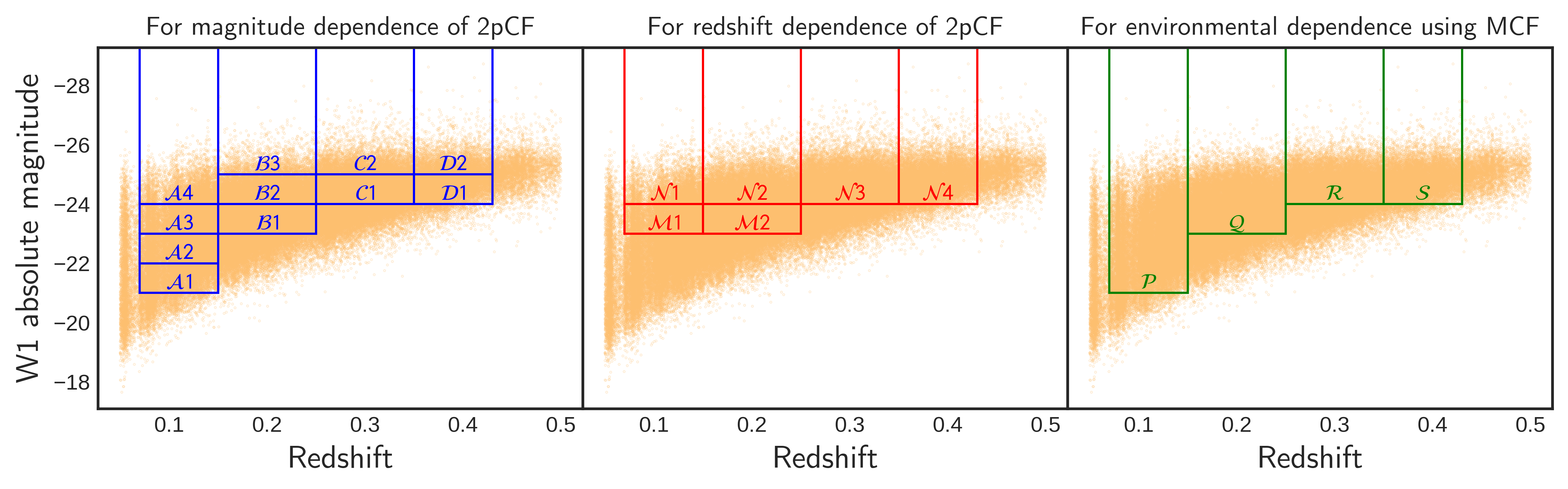}
    \caption{Selection of subsamples used to study the W1 absolute magnitude dependence of 2pCF (left panel), redshift dependence of 2pCF (middle panel), and environmental dependence of galaxy properties using MCF (right panel).}
    \label{fig:subsamples}
\end{figure*}

To study the W1 absolute magnitude dependence of 2pCF, we first binned the galaxies into four different redshift bins: $\mathcal{A}$ ($0.07 \leq z < 0.15$), $\mathcal{B}$ ($0.15 \leq z < 0.25$), $\mathcal{C}$ ($0.25 \leq z < 0.35$), and $\mathcal{D}$ ($0.35 \leq z < 0.43$).
Each redshift bin was further divided into different W1 absolute magnitude bins, giving a total of 11 samples (named from $\mathcal{A}$1 to $\mathcal{D}$4).
This selection is shown in the left panel of Fig.~\ref{fig:subsamples}, and the details such as sample name, redshift range, W1 absolute magnitude range, and number of galaxies are given in Table~\ref{table:subsamples_w1dep}.

\begin{table}
\caption{Definitions and the number of galaxies of different galaxy subsamples used to study the W1 absolute magnitude dependence of 2pCF.}
\begin{center}
\begin{tabular}{l c c c}
\toprule
\toprule
  \multicolumn{1}{l}{Sample} &
  \multicolumn{1}{c}{$z^\text{range}$} &
  \multicolumn{1}{c}{$M_\text{W1}^\text{range}$} &
  \multicolumn{1}{c}{$N_\text{gal}$}
  \\ 
\midrule
  $\mathcal{A}$1 & \multirow{4}{*}{$0.07 \leq z < 0.15$} & $-22 < M_\text{W1} \leq -21$ & 7487  \\
  $\mathcal{A}$2 & & $-23 < M_\text{W1} \leq -22$ & 10758  \\
  $\mathcal{A}$3 & & $-24 < M_\text{W1} \leq -23$ & 9945  \\
  $\mathcal{A}$4 & & $M_\text{W1} \leq -24$ & 5040  \\
  \midrule
  $\mathcal{B}$1 & \multirow{3}{*}{$0.15 \leq z < 0.25$} & $-24 < M_\text{W1} \leq -23$ & 23521 \\
  $\mathcal{B}$2 & & $-25 < M_\text{W1} \leq -24 $ & 14657  \\
  $\mathcal{B}$3 & & $M_\text{W1} \leq -25 $ & 2267 \\
  \midrule
  $\mathcal{C}$1 & \multirow{2}{*}{$0.25 \leq z < 0.35$} & $-25 < M_\text{W1} \leq -24 $ & 26608  \\
  $\mathcal{C}$2 & & $M_\text{W1} \leq -25 $ & 5704 \\
  \midrule
  $\mathcal{D}$1 & \multirow{2}{*}{$0.35 \leq z < 0.43$} & $-25 < M_\text{W1} \leq -24 $ & 10191  \\
  $\mathcal{D}$2 & & $M_\text{W1} \leq -25 $ & 4770 \\
\bottomrule
\end{tabular}
\end{center}
\label{table:subsamples_w1dep}
\end{table}

To study the redshift dependence of 2pCF, we need subsamples with varying redshift in the same magnitude range.
To obtain them, we first created two magnitude bins $\mathcal{M}$ ($-24 < M_\text{W1} \leq -23$) and $\mathcal{N}$ ($M_\text{W1} \leq -24$).
Each magnitude bin was further divided into different redshift bins, giving a total of six subsamples (called $\mathcal{M}$1 to $\mathcal{N}$4), as shown in the middle panel of Fig.~\ref{fig:subsamples}.
The properties of these subsamples such as sample name, W1 absolute magnitude range, redshift range, and number of galaxies are given in Table~\ref{table:subsamples_zdep}.

Although we define a total of 17 samples in Table~\ref{table:subsamples_w1dep} and Table~~\ref{table:subsamples_zdep}, some of the samples overlap ($\mathcal{A}$3-$\mathcal{M}$1, $\mathcal{A}$4-$\mathcal{N}$1, and $\mathcal{B}$1-$\mathcal{M}$2), leaving 14 unique samples.
However, we named them differently for clarity and easier interpretation.

\begin{table}
\caption{Definitions of the galaxy subsamples used to study the redshift dependence of 2pCF.}
\begin{center}
\begin{tabular}{l c c c}
\toprule
\toprule
  \multicolumn{1}{l}{Sample} &
  \multicolumn{1}{c}{$M_\text{W1}^\text{range}$} &
  \multicolumn{1}{c}{$z^\text{range}$} &
  \multicolumn{1}{c}{$N_\text{gal}$}
  \\ 
\midrule 
  $\mathcal{M}$1 & \multirow{2}{*}{$-24 < M_\text{W1} \leq -23$} & $0.07 \leq z < 0.15$ & 9945  \\
  $\mathcal{M}$2 & & $0.15 \leq z < 0.25$ & 23521  \\
  \midrule
  $\mathcal{N}$1 & \multirow{4}{*}{$M_\text{W1} \leq -24$} & $0.07 \leq z < 0.15$ & 5040 \\
   $\mathcal{N}$2 & & $0.15 \leq z < 0.25$ & 16924  \\
   $\mathcal{N}$3 & & $0.25 \leq z < 0.35$ & 32312 \\
   $\mathcal{N}$4 & & $0.35 \leq z < 0.43$ & 14961 \\
\bottomrule
\end{tabular}
\end{center}
\label{table:subsamples_zdep}
\end{table}

\subsection{W2 to W4 selection of the $r$+W1 magnitude-limited samples}\label{sec:data_w1w4selection}

By a cross-matching technique using  \texttt{CATAID}, we selected the galaxies in the parent sample with matches in the GAMA-WISE and ProSpect catalogues. 
Then we assigned {\small $L_\text{W1}$}, {\small $L_\text{W2}$}, {\small $L_\text{W3}$}, {\small $L_\text{W4}$}, {\small $M^{\star}_\text{WISE}$}, {\small $\text{SFR12}_\text{WISE}$}, and {\small $\text{SFR22}_\text{WISE}$} from the GAMA-WISE catalogue and {\small $M^{\star}_\text{ProSpect}$} and  {\small $\text{SFR}_\text{ProSpect}$} from the ProSpect catalogue.
As demonstrated in the right panel of Fig.~\ref{fig:subsamples}, we selected only the galaxies above a certain W1 absolute magnitude limit in each redshift bin.
This returned four $r$+W1 selected samples: $\mathcal{P}$, $\mathcal{Q}$, $\mathcal{R}$, and $\mathcal{S}$.

Furthermore, in each of these four samples, we applied three additional selections based on their luminosities in the W2, W3, and W4 bands, resulting in a total number of 16 subsamples.
Specifically, in this context, we defined a selection in a particular band as selecting only those galaxies with available luminosity measurements in the respective band.
That is, the W2 selection selects only galaxies with available W2 luminosity, W3 selection selects only galaxies with available W3 luminosity, and so on.
For example, sample $\mathcal{P}$ contains 31555, 30991, 25862, and 16799 galaxies with W1, W2, W3, and W4 luminosities, respectively.
It is to be noted that the first selection was made with the W1 absolute magnitude, and then in each following selection, the galaxies without longer wavelength luminosities were filtered out. 
Therefore, the resulting samples are with selections $r$+W1, $r$+W1+W2, $r$+W1+W3, and $r$+W1+W4.
The differences between these selections are due to the difference in sensitivity of WISE bands, with W3 and W4 being less sensitive than W1 and W2 \citep{wright2010_wise, jarrett2013}.

The properties of these subsamples such as sample name, redshift range, upper limiting W1 absolute magnitude, and number of galaxies in each selections are given in Table~\ref{table:subsamples_mcf}.
For each of the 16 subsamples, we measured the 2pCFs and MCFs using the properties with available measurements in that particular selection.

\begin{figure*}[t]
    \centering
    \includegraphics[width=0.79\linewidth]{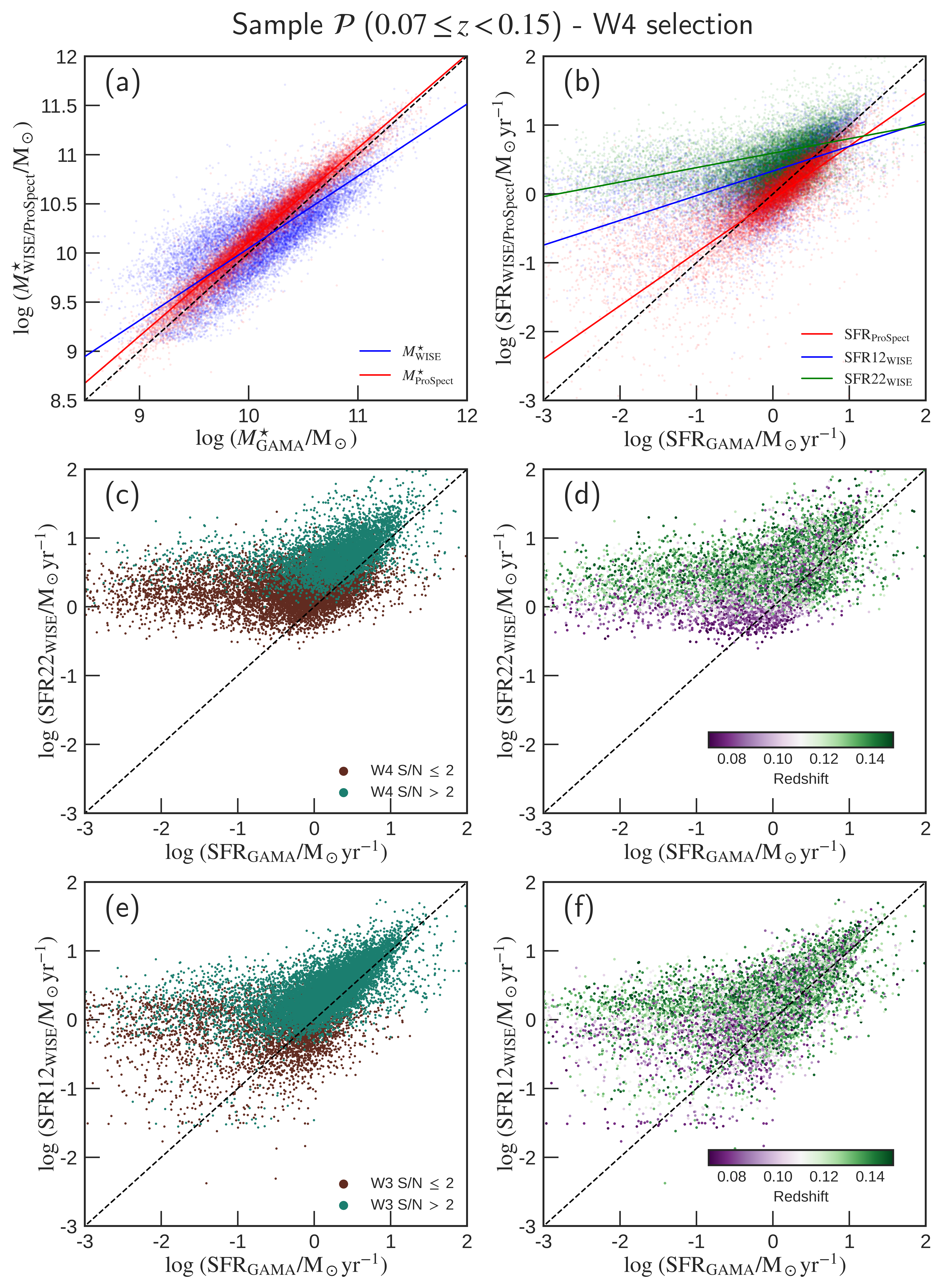}
    \caption{
    Comparison of the properties estimated using different techniques of the galaxies in the W4 selection of sample $\mathcal{P}$.
    (a) WISE and ProSpect estimates of stellar mass plotted against the GAMA stellar masses. 
    (b)~WISE and ProSpect estimates of SFR plotted against the GAMA SFRs. 
    (c)~{\small $\text{SFR}_\text{GAMA}$}--{\small $\text{SFR22}_\text{WISE}$} distribution colour-coded with S/N values.
    (d)~{\small $\text{SFR}_\text{GAMA}$}--{\small $\text{SFR22}_\text{WISE}$} distribution colour-coded with redshift.
    (e)~{\small $\text{SFR}_\text{GAMA}$}--{\small $\text{SFR12}_\text{WISE}$} distribution colour-coded with S/N values.
    (f)~{\small $\text{SFR}_\text{GAMA}$}--{\small $\text{SFR12}_\text{WISE}$} distribution colour-coded with redshift.
    The solid coloured lines in panels (a) and (b) show the corresponding linear fit function of the population, as labelled.
    The dashed black lines in all the panels show the 1:1 correlation line.
    }
    \label{fig:M*-sft-scatter}
\end{figure*}

In Fig.~\ref{fig:M*-sft-scatter} we compare the properties estimated using different techniques of the galaxies in the W4 selection of sample $\mathcal{P}$.
Panel~(a) contains a comparison of {\small $M^{\star}_\text{WISE}$} and {\small $M^{\star}_\text{ProSpect}$} with {\small $M^{\star}_\text{GAMA}$}.
Panel~(b) shows a comparison of {\small $\text{SFR12}_\text{WISE}$}, {\small $\text{SFR22}_\text{WISE}$}, and {\small $\text{SFR}_\text{ProSpect}$} against {\small $\text{SFR}_\text{GAMA}$}. 
For stellar mass and SFR, we observe deviations from the identity line.
In Sect.~\ref{sec:discussion_diffest} we discuss whether the differences in the estimates from different techniques influence the galaxy clustering measurements.

Fig.~\ref{fig:M*-sft-scatter} shows that the WISE properties scatter significantly.
As mentioned in Sect.~\ref{data_gamawise}, our selected samples contain galaxies with low S/N in the WISE bands.
For instance, out of the galaxies in the W4 selection of sample $\mathcal{P}$, almost all have S/N>2 in W1 band, $\sim$93\% have S/N>2 in W2 band, $\sim$70\% have S/N>2 in W3 band, and $\sim$32\% have S/N>2 in W4 band. 
Their low S/N counterparts having low-confidence flux measurements might be the reason for the scatter we observe. 
For example, in panel~(c) of Fig.~\ref{fig:M*-sft-scatter}, we show the {\small $\text{SFR}_\text{GAMA}$}~--~{\small $\text{SFR22}_\text{WISE}$} distribution with galaxies with S/N$\leq$2 in W4 band marked in brown.
The majority of the galaxies that are away from the unity line clearly have S/N$\leq$2.
In panel~(e), we show a similar plot in the case of {\small $\text{SFR12}_\text{WISE}$}.
In panels~(d) and (f), we colour-code the same distribution with the redshift.
For $z>0.1$ sources, the W3 and W4 band detections are affected and hence bias the distribution \citep{cluver2020}.
However, we do not expect these scatters to significantly bias our clustering measurements because our MCF measurements are based on the rank of the galaxy rather than on the value itself (see Sect.~\ref{sec:measurement_mcf}).

\begin{table}[h]
\caption{Definitions of the galaxy subsamples used to study the environmental dependence of galaxy properties using MCFs.}
\begin{center}
{\tiny
\begin{tabular}{l c c c c c c}
\toprule
\toprule
  \multirow{2}{*}{Sample} &
  \multirow{2}{*}{$z^\text{range}$} &
  \multirow{2}{*}{$M_\text{W1}^\text{lim}$} &
  \multicolumn{4}{c}{$N_\text{gal}^\text{selection}$}
  \\ 
  \cmidrule{4-7}
  & & (upper) & W1 & W2 & W3 & W4 \\
\midrule 
  $\mathcal{P}$ & $0.07 \leq z < 0.15$ & $-21$ & 31555 & 30991 & 25862 & 16799  \\
  $\mathcal{Q}$ & $0.15 \leq z < 0.25$ & $-23$ & 38135 & 38115 & 30529 & 19877 \\
  $\mathcal{R}$ & $0.25 \leq z < 0.35$ & $-24$ & 30265 & 30263 & 23034 & 15023 \\
  $\mathcal{S}$ & $0.35 \leq z < 0.43$ & $-24$ & 14224 & 14222 & 10564 & 6447 \\
\bottomrule
\end{tabular}
}
\end{center}
\label{table:subsamples_mcf}
\end{table}

\subsection{Random samples}\label{sec:data_random}

To estimate 2pCF of a galaxy sample, we require a random sample that reflects the redshift and sky distribution of real sample.
To generate random samples, we used the GAMA random galaxy catalogue (\textsc{Randomsv02} DMU) by \citet{farrow2015_gama_cf}.
They used the method of \citet{cole2011} that generates clones of real galaxies, where the number of clones generated for a given galaxy is proportional to the ratio of the maximum volume out to which that galaxy is visible given the $r$-band flux limit ($V_\text{max, \textit{r}}$) to the same volume weighted by the number density.
The generation of clones also accounts for targeting and redshift incompleteness.
This DMU provides around 400 random galaxies per real GAMA galaxy, and these random galaxies share the \texttt{CATAID} and all the physical properties of the real galaxy. 
Using the \texttt{CATAID}, we assigned the W1 absolute magnitude to each random galaxy.
Then for all the selected samples, random samples were created by applying the corresponding redshift and magnitude limits as given in Tables~\ref{table:subsamples_w1dep}, \ref{table:subsamples_zdep}, and \ref{table:subsamples_mcf} and randomly selecting 5-10 times the number of galaxies in the corresponding real samples.

It is to be noted that only the $r$-band selection is considered in the generation of clones, that is, the clones of a galaxy are distributed within its $V_\text{max, \textit{r}}$.
However, our selection contains an additional W1 magnitude selection.
This difference in selection might affect the redshift distribution of the clones.
To solve this problem, we adopted the weighting scheme proposed by \citet{gunawardhana2018}.
For a given galaxy $i$ in the real sample, we defined a weight for all its clones, given by

\begin{equation}\label{eqn:randomWeight}
    w^i = \frac{N^i_{V_\text{max, \textit{r}}}}{N^i_{\text{min}(V_\text{max, W1}, V_\text{max, \textit{r}}, V_\text{zlim})}}
,\end{equation}
where $N^i_{V_\text{max, \textit{r}}}$ is the total number of clones of the galaxy $i$ that are distributed randomly within $V_\text{max, \textit{r}}$, and $N^i_{\text{min}(V_\text{max, W1}, V_\text{max, \textit{r}}, V_\text{zlim})}$ is the number of clones within $\text{min}(V_\text{max, W1}, V_\text{max, \textit{r}}, V_\text{zlim})$.
Here, $V_\text{max, \textit{r}}$ was computed using $z_\text{max}$ value at $r=19.8$ taken from \textsc{StellarMassesLambdarv20} DMU \citep{taylor2011_gama_stellar_mass}, $V_\text{max, W1}$ was computed using $z_\text{max}$ at W1 limiting magnitude W1(AB)=$19.3$ \citep{jarrett2017}, and $V_\text{zlim}$ corresponds to the maximum redshift limit of the sample. 
After applying all these selections and weights for each sample, we confirm the agreement between $N(z)$ of the real and random samples.

\section{Measurement methods}\label{sec:measurement}

We used galaxy 2pCF and MCF to quantify galaxy clustering and its dependence on environment. 
As this work is an extension of \citet{sureshkumar2021}, all methods used in this study are  the same as used therein. 
Therefore, we refer to Sect.~3 of that paper for detailed description, while we provide a short summary here.

\subsection{Galaxy two-point correlation function}\label{sec:measurement_2pcf}

The galaxy 2pCF, $\xi(r)$ quantifies the clustering of galaxies. 
It is defined as the excess probability of observing a pair of galaxies separated by a given $r$ in a volume element $dV$ \citep{peebles1980} over a random distribution,

\begin{equation}\label{eqn:dp}
    \mathrm dP = \left( n \, \mathrm dV \right)^2 \, [ 1 + \xi(r) ] ,
\end{equation}

where $n$ is the number density of galaxies.

To estimate 2pCF using the real and random samples, we calculated the separation between all the real-real, random-random, and real-random galaxies using their sky positions and redshift.
Then we counted the number of galaxy pairs in different separation bins, and the 2pCF was computed using these pair counts.

Several effects need to be taken into account in 2pCF measurements, particularly, galaxy peculiar motions. 
In addition to the motion due to the expansion of space, the galaxies possess peculiar velocities caused by the local gravitational field. 
Therefore, the positions of galaxies in redshift space (where the observed redshift is used as a proxy for distance) differ from those in real space (where the physical distance is considered).
This effect, known as redshift-space distortion (RSD), leads to the fingers-of-god effect on smaller scales and to the Kaiser effect on larger scales \citep{kaiser1987}.
It affects the 2pCF measurements in redshift space \citep{davis&peebles1983}.

As a standard practise to account for this distortion, we decomposed the pair separation into components perpendicular ($r_\text{p}$) and parallel ($\pi$) to the line of sight.
We then estimated the two-dimensional 2pCF $\xi(r_\text{p},\pi)$ using \citet{landy&szalay1993} estimator given by

\begin{equation} \label{eqn:landy-szalay} 
    \xi(r_\text{p},\pi)= \frac{\langle DD(r_\text{p},\pi) \rangle - 2 \langle DR(r_\text{p},\pi) \rangle + \langle RR(r_\text{p},\pi) \rangle}{\langle RR(r_\text{p},\pi) \rangle} ,
\end{equation}

where $\langle DD \rangle$, $\langle DR \rangle$ and $\langle RR \rangle$ are the normalised real-real, real-random, and random-random pair counts in the corresponding $(r_\text{p},\pi)$ bin.
While computing $DR$ and $RR$, each random galaxy carries the weight measured using Eq.~\ref{eqn:randomWeight}.

We then integrated $\xi(r_\mathrm{p},\pi)$ over the line-of-sight ($\pi$) direction to obtain the projected 2pCF $\omega_\mathrm{p}(r_\mathrm{p})$, which can be used to recover the real-space 2pCF devoid of RSD \citep{davis&peebles1983}.
The projected 2pCF is defined as

\begin{equation}\label{eqn:projectedcf}
    \omega_\mathrm{p}(r_\mathrm{p}) = 2 \, \int_0^{\pi_\text{max}} \xi(r_\mathrm{p},\pi) \, \mathrm d\pi 
,\end{equation}

where the factor of 2 comes from the fact that $\xi(r_\mathrm{p},\pi)$ is a symmetric function along $r_\mathrm{p}$.

Following Appendix B of \citet{loveday2018_gama_pvd}, we chose the limit of integration $\pi_\text{max}$ to be $40 \, h^{-1} \mathrm{Mpc}$ , which is reasonable enough to include all the correlated pairs and reduce the noise in the estimator. 

\subsection{Error estimate of 2pCF}\label{sec:measurement_errors_2pcf}

To estimate the errors in 2pCF, we used the jackknife resampling method \citep{norberg2009} with nine jackknife regions.
The associated covariance matrix is given by

\begin{equation}\label{eqn:covmat_jackknife}
C_{ij} = \frac{N_\text{JK} - 1}{N_\text{JK}} \sum\limits_{k=1}^{N_\text{JK}} \left( \omega_\mathrm{p}^k(r_i) - \bar\omega_\mathrm{p}(r_i) \right) \, \, \left( \omega_\mathrm{p}^k(r_j) - \bar\omega_\mathrm{p}(r_j) \right) ,
\end{equation}

where $\omega_\mathrm{p}^k(r_j)$ is the $\omega_\mathrm{p}$ value at $r_\mathrm{p} = r_j$ in the $k$th jackknife copy, and $\bar{\omega_\mathrm{p}}$ is the average 2pCF from $N_\text{JK}$ copies. 
The error bar for the $\omega_\mathrm{p}$ at the $i$th bin comes from the square root of the diagonal element ($\sqrt{C_{ii}}$) of the covariance matrix.

\subsection{Power-law fits of 2pCF}\label{sec:measurement_powerlaw}

It has been observed that 2pCF in the intermediate separation scale follows a power law \citep{groth&peebles1977} given by

\begin{equation}\label{eqn:powerlaw}
\xi(r)=\left(\frac{r}{r_0}\right)^{-\gamma} ,
\end{equation}

where the power-law fit parameters $r_0$ and $\gamma$ are the correlation length and slope, respectively.
In the power-law approximation, the strength of the clustering of a sample can be quantified using these two parameters.
A higher value of $r_0$ signifies a larger amplitude of the correlation function, and the higher value of $\gamma$ signifies stronger clustering at smaller scales.

We estimated the power-law fit parameters $r_0$ and $\gamma$ using the covariance matrix given in Eq.~\ref{eqn:covmat_jackknife}.
Using the inverse of correlation matrix $\tilde{C}^{-1}$ (normalised covariance matrix), we minimised $\chi^2$ given by,

\begin{equation}\label{eqn:chi_square}
 \chi^2 = \sum \limits_{i,j} \frac{\left( \omega_\mathrm{p}^\text{mod}(r_i) - \omega_\mathrm{p}(r_i) \right)}{\sigma_i} \,\, \tilde{C}^{-1}_{ij} \,\, \frac{\left( \omega_\mathrm{p}^\text{mod}(r_j) - \omega_\mathrm{p}(r_j) \right)}{\sigma_j} ,
\end{equation}

where $\sigma_i = \sqrt{C_{ii}}$ and $\omega_\mathrm{p}^\text{mod}$ is the power-law model value given by

\begin{equation}\label{eqn:analytic_wp}
\omega_\mathrm{p}(r_\mathrm{p}) = r_\mathrm{p} \, \left( \frac{r_\mathrm{p}}{r_0} \right)^{-\gamma}  \, \frac{\Gamma \left(\frac{1}{2}\right) \, \Gamma \left(\frac{\gamma-1}{2} \right)}{\Gamma \left(\frac{\gamma}{2}\right)} ,
\end{equation}

where $\Gamma(n)$ is Euler's gamma function \citep{davis&peebles1983}.

As described in Sect.~3.4 of \citet{sureshkumar2021}, we used the singular-value decomposition method to estimate $\tilde{C}^{-1}$.
This efficiently removed the influence of noisy eigenmodes of the covariance matrix and provided reliable power-law fit parameters.
Detailed descriptions of the fitting procedure using singular-value decomposition to estimate the power-law parameters are given in \citet{gaztanaga2005} and \citet{marin2013}.

\subsection{Marked correlation function}\label{sec:measurement_mcf}

The marked correlation function is a statistical tool for studying the environmental dependence of galaxy properties by assigning a mark to each galaxy \citep{skibba2013}.
The mark can be any galaxy property such as luminosity, colour, stellar mass, SFR, and morphology \citep{sheth2005_galform_models}.

The two-point MCF in real space is defined as

\begin{equation}\label{eqn:M(r)}
M(r) = \frac{1 + W(r)}{1 + \xi(r)},
\end{equation}

where $\xi(r)$ is the galaxy 2pCF and $W(r)$ is the weighted 2pCF.
$W(r)$ was computed using the same estimator as $\xi(r)$, 
but with each real galaxy weighted by a ratio of its mark to the mean mark value of the sample.
For example, in case of stellar mass MCF, we used $\text{weight} = \text{mass} / \text{mean mass}$.

Similar as in the case of 2pCF, RSD can also affect the MCF.
We therefore adopted the same method of measuring the projected two-point MCF to account for this effect.
It is then defined as

\begin{equation}\label{eqn:projectedMCF}
    M_\mathrm{p}(r_\mathrm{p}) = \frac{1 + W_\mathrm{p}(r_\mathrm{p})/r_\mathrm{p}}{1 + \omega_\mathrm{p}(r_\mathrm{p})/r_\mathrm{p}} ,
\end{equation}

where $W_\mathrm{p}(r_\mathrm{p})$ is the projected weighted 2pCF estimated using Eq.~\ref{eqn:landy-szalay} and then Eq.~\ref{eqn:projectedcf}, but with weighted real galaxy pairs.
In Eq.~\ref{eqn:projectedMCF}, $W_\mathrm{p}$ and $\omega_\mathrm{p}$ are divided by $r_\mathrm{p}$ because they both have a dimension of length and $M_\mathrm{p}$ is dimensionless.

For any given property, $\mathrm{MCF}=1$ shows a lack of correlation between that property and the environment.
The strength of the deviation from unity signifies the strength of the correlation ($\mathrm{MCF} > 1$) or the anti-correlation ($\mathrm{MCF} < 1$) between the property and the environment. 

Different galaxy properties have different ranges and are expressed in different scales (log or linear). 
Therefore, a comparison of MCFs obtained using the direct value of the property as a mark would be unreliable, if not impossible. 
To solve this problem, we used the rank-ordered mark correlation function proposed by \citet{skibba2013}. 
The method of computing the MCF remained the same, but instead of using the galaxy property as a mark, we assigned a rank to each property value, that is, the galaxy with the greatest value of the property carries the highest rank.
Then we used these ranks to measure the MCF. 
This makes the MCFs measured using different properties comparable to each other.
In this work, we present the rank-ordered MCFs \citep[similarly as in ][]{sureshkumar2021}.

\subsection{Error estimate of the MCF}\label{sec:measurement_errors_mcf}

There are two major uncertainties in MCFs. 
They are related to the role of spatial positions and marks of the galaxies.
The spatial distribution uncertainty can be estimated using the jackknife method as described in Sect.~\ref{sec:measurement_errors_2pcf}.
The mark uncertainty is related to the correlation between marks and galaxy positions.
This uncertainty can be estimated from the variance over redistribution of the marks randomly among galaxies. 
We randomly shuffled the marks among the galaxies in the sample and then remeasured the MCF. 
The standard deviation over a number of MCF measurements with such shuffled marks ($\sim$ 100 times) gives the uncertainty in the MCF \citep{beisbart2000, skibba2006}.
As both of these errors are important, we present a combined error calculated using the summation in quadrature method.

\section{Results}\label{sec:results}

In this section, we present the results of three different analyses: the W1 absolute magnitude dependence of 2pCF, the redshift dependence of 2pCF, and the environmental dependence of various galaxy properties using MCF. 
The MCFs were measured using the luminosities in WISE bands, stellar masses, and SFRs computed by \citet{taylor2011_gama_stellar_mass}, \citet{cluver2014, cluver2017}, and \citet{robotham2020}.
All the 2pCFs were fitted with power-law models using the method described in Sect.~\ref{sec:measurement_powerlaw}.
The errors of 2pCFs were obtained from nine jackknife realisations, and those of MCFs were obtained by combining the errors from the jackknife method and the random mark shuffling method as described in Sect.~\ref{sec:measurement_errors_mcf}.

\subsection{Dependence of 2pCF on W1 magnitude}\label{sec:result_w1dep}

\begin{figure*}[h]
    \centering
    \includegraphics[width=\linewidth]{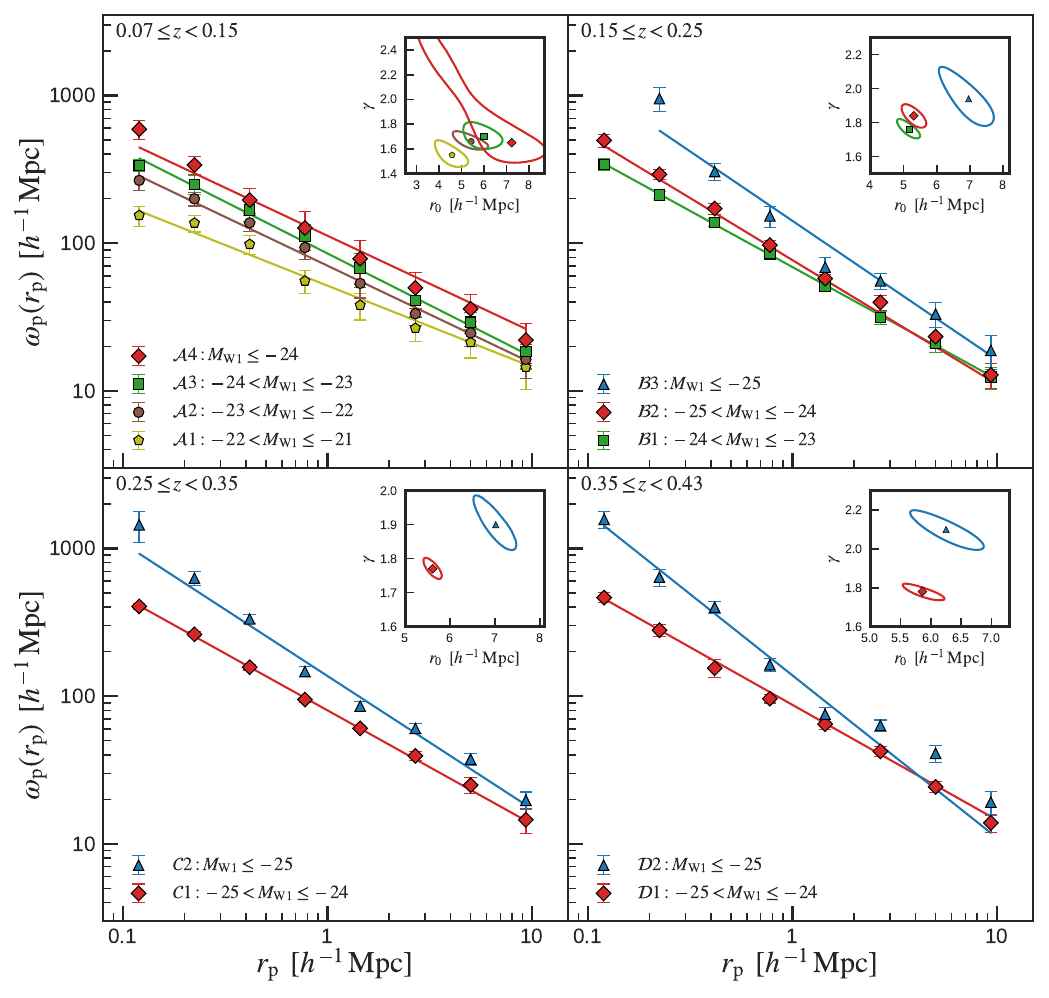}
    \caption{Projected two-point correlation functions corresponding to different W1 absolute magnitude bins (as labelled) in four different redshift bins: $0.07 \leq z < 0.15$ (\textit{upper left panel}), $0.15 \leq z < 0.25$ (\textit{upper right panel}), $0.25 \leq z < 0.35$ (\textit{lower left panel}), and $0.35 \leq z < 0.43$ (\textit{lower right panel}).
    The insets show the best-fitting power-law parameters with the $1\sigma$ error contour.
    The error bars on the markers are the square root of the diagonals of the covariance matrix obtained from the jackknife resampling method.}
    \label{fig:result_w1dep}
\end{figure*}

In Fig.~\ref{fig:result_w1dep} we show the measurements of 2pCF with their corresponding power-law fits for the 11 samples described in Table~\ref{table:subsamples_w1dep}.
The four different panels of Fig.~\ref{fig:result_w1dep} show the measurements in four different redshift bins, and in each panel, we show the 2pCF for samples with varying W1 absolute magnitude cuts, as labelled (the markers and lines of samples representing similar magnitude cuts have similar colours in all the panels).
In the insets of each panel, we show the best-fitting power-law parameters with corresponding $1\sigma$ error contours.
The parameters are also listed in Table~\ref{table:wp_params_w1dep}.

\begin{table}[t]
\caption{Best-fitting power-law parameters for the galaxy samples we used to study the W1 absolute magnitude dependence of 2pCF.}
\begin{center}
\begin{tabular}{l c c c}
\toprule
\toprule
  \multicolumn{1}{l}{Sample} &
  \multicolumn{1}{c}{$r_0 \, (h^{-1} \mathrm{Mpc})$} &
  \multicolumn{1}{c}{$\gamma$} &
  \multicolumn{1}{c}{$\chi^2$}
  \\ 
\midrule
  $\mathcal{A}$1 & $4.58 \pm 0.77$ & $1.55 \pm 0.12$ & 0.01 \\ 
  $\mathcal{A}$2 & $5.43 \pm 0.82$ & $1.66 \pm 0.08$ & 0.29 \\ 
  $\mathcal{A}$3 & $5.99 \pm 0.91$ & $1.70 \pm 0.12$ & 0.01 \\ 
  $\mathcal{A}$4 & $7.23 \pm 4.14$ & $1.65 \pm 0.84$ & 0.01 \\ 
  \midrule
  $\mathcal{B}$1 & $5.17 \pm 0.37$ & $1.76 \pm 0.06$ & 0.12 \\ 
   $\mathcal{B}$2 & $5.31 \pm 0.37$ & $1.84 \pm 0.07$ & 1.71 \\ 
   $\mathcal{B}$3 & $6.96 \pm 0.89$ & $1.94 \pm 0.19$ & 3.52 \\ 
  \midrule
  $\mathcal{C}$1 & $5.61 \pm 0.21$ & $1.77 \pm 0.03$ & 1.71 \\ 
  $\mathcal{C}$2 & $7.01 \pm 0.50$ & $1.90 \pm 0.09$ & 5.94 \\ 
  \midrule
 $\mathcal{D}$1 & $5.86 \pm 0.36$ & $1.78 \pm 0.04$ & 1.38 \\ 
  $\mathcal{D}$2 & $6.25 \pm 0.62$ & $2.10 \pm 0.10$ & 15.82 \\ 
\bottomrule
\end{tabular}
\end{center}
\label{table:wp_params_w1dep}
\end{table}

\begin{figure}[h]
    \centering
    \includegraphics[width=\linewidth]{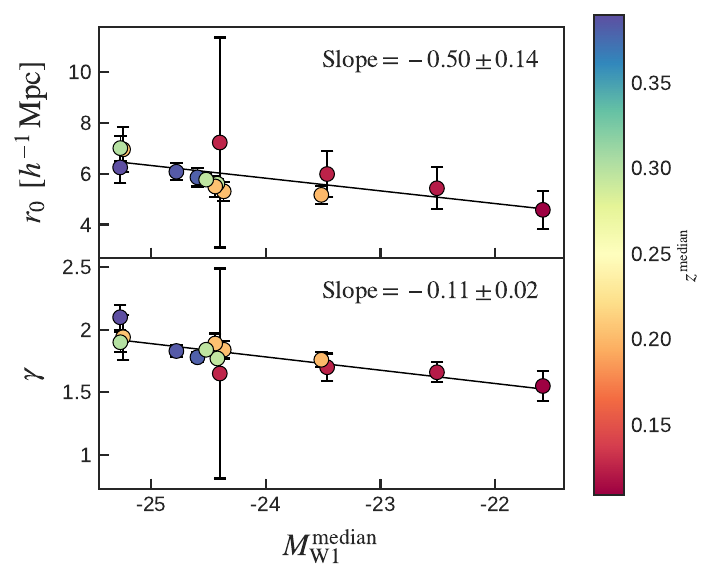}
    \caption{Dependence of the power-law fit parameters $r_0$ and $\gamma$ on the median W1 absolute magnitude colour-coded with the median redshift of the 14 unique samples described in Sect.~\ref{sec:data_sampleselection_2pcf}.
    The black line in both panels represents the corresponding best linear fits whose slopes are marked accordingly.
    }
    \label{fig:result_w1zr0gam}
\end{figure}

In Fig.~\ref{fig:result_w1zr0gam} we show the dependence of the power-law fit parameters of 2pCF on the median W1 absolute magnitude. 
We observe a decrease in $r_0$ and $\gamma$ with an increase in W1 absolute magnitude (i.e. decrease in the W1 luminosity).
To quantify this observation, we fit a linear function to the $M_\mathrm{W1}^{\mathrm{median}}-r_0$ and $M_\mathrm{W1}^{\mathrm{median}}-\gamma$ relations.
We find that the slope in the case of $r_0$ is $-0.50 \pm 0.14$ and that in the case of $\gamma$ is $-0.11 \pm 0.02$.
These slopes are non-vanishing with a significance of $3.5\sigma$ and $5.5\sigma,$ respectively.
This implies a strong dependence of the galaxy clustering on the W1 luminosity, that is, W1 bright galaxies cluster more strongly than faint ones.
This is inline with the observation of \citet{jarrett2017}.

\subsection{Dependence of 2pCF on redshift}\label{sec:result_zdep}

\begin{figure*}[h]
    \centering
    \includegraphics[width=\linewidth]{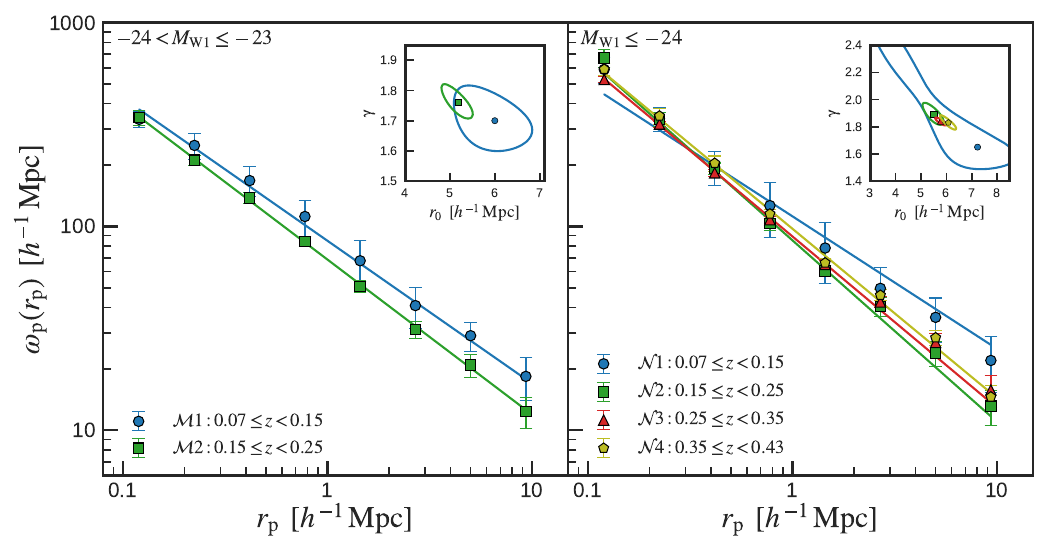}
    \caption{Projected two-point correlation functions corresponding to different redshift bins (as labelled) in the magnitude ranges $-24 < M_\text{W1} \leq -23$ (\textit{left panel}) and $M_\text{W1} \leq -24$ (\textit{right panel}).
    The insets show the best fitting power-law parameters with $1\sigma$ error contour.
    The error bars on the markers are the square root of the diagonals of the covariance matrix obtained from the jackknife resampling method.}
    \label{fig:result_zdep}
\end{figure*}

In Fig.~\ref{fig:result_zdep} we show the 2pCF measurements of the six samples described in Table~\ref{table:subsamples_zdep}.
The left panel shows the measurements at different redshift ranges for the fainter galaxies in W1, and the right panel shows those for the brighter galaxies.
The same coloured curves in the panels represent the same redshift ranges.
The best-fitting power-law parameters for these samples are given in Table~\ref{table:wp_params_zdep}.

\begin{table}[h]
\caption{Best-fitting power-law parameters for the galaxy samples we used to study the redshift dependence of 2pCF.}
\begin{center}
\begin{tabular}{l c c c}
\toprule
\toprule
  \multicolumn{1}{l}{Sample} &
  \multicolumn{1}{c}{$r_0 \, (h^{-1} \mathrm{Mpc})$} &
  \multicolumn{1}{c}{$\gamma$} &
  \multicolumn{1}{c}{$\chi^2$}
  \\ 
\midrule 
  $\mathcal{M}$1 & $5.99 \pm 0.91$ & $1.70 \pm 0.12$ & 0.01 \\ 
  $\mathcal{M}$2 & $5.17 \pm 0.37$ & $1.76 \pm 0.06$ & 0.12 \\ 
  \midrule
  $\mathcal{N}$1 & $7.23 \pm 4.14$ & $1.65 \pm 0.84$ & 0.01 \\ 
   $\mathcal{N}$2 & $5.50 \pm 0.47$ & $1.89 \pm 0.09$ & 3.27 \\ 
   $\mathcal{N}$3 & $5.77 \pm 0.22$ & $1.84 \pm 0.03$ & 2.15 \\ 
  $\mathcal{N}$4 & $6.08 \pm 0.36$ & $1.83 \pm 0.05$ & 2.18 \\ 
\bottomrule
\end{tabular}
\end{center}
\label{table:wp_params_zdep}
\end{table}

We studied the redshift evolution of the correlation function of galaxy samples that were equally limited by the W1 absolute magnitude. 
From the best-fitting power-law parameters, it is observed that the clustering strength does not significantly vary between galaxy samples at different redshifts.
The correlation lengths of $\mathcal{M}$1 and $\mathcal{M}$2 are comparable within $1\sigma$.
The same is the case with $\mathcal{N}$2 and $\mathcal{N}$4. 
We omit sample $\mathcal{N}$1 from this comparison because of its unreliable measurements of power-law fit parameters with large uncertainties.

It is also to be noted that the faintest galaxy samples are prone to incompleteness due to the survey flux-limit. 
That is, those samples may miss W1-fainter galaxies near to the flux limit. 
These missing galaxies can give an apparent rise in the correlation length.
This means that the correlation lengths of the complete samples might be shorter than the current measurements. 
However, this shift in $r_0$ would only strengthen the  $M_\mathrm{W1}^{\mathrm{median}}-r_0$ slope in Fig.~\ref{fig:result_w1zr0gam} and hence does not affect the conclusions.

\subsection{Dependence of 2pCF on W1 to W4 selection}\label{sec:result_2pcf_w1tow4}

\begin{figure*}[h]
    \centering
    \includegraphics[width=0.9\linewidth]{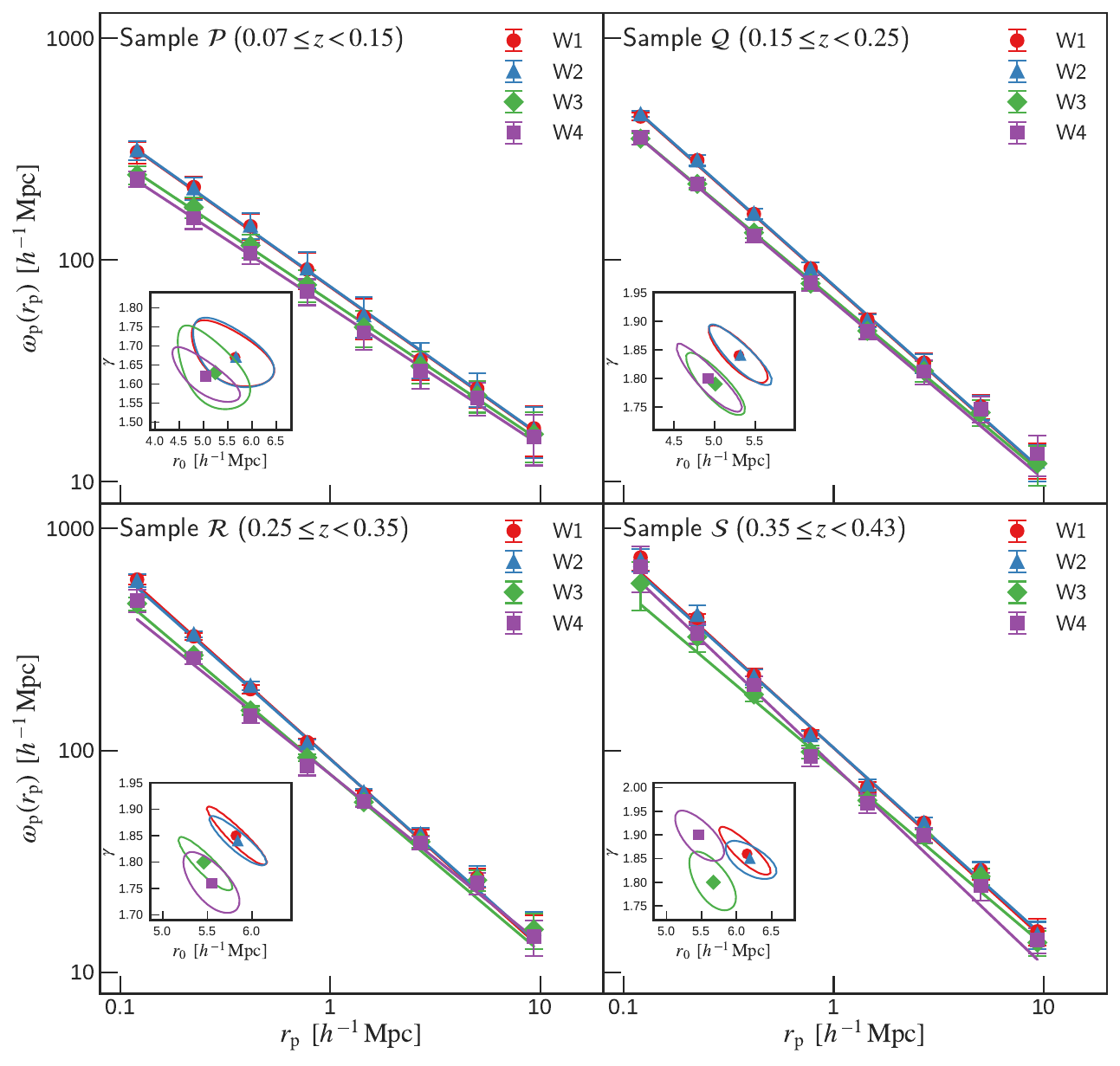}
    \caption{Projected two-point correlation functions corresponding to the W1, W2, W3, and W4 selections applied in the samples $\mathcal{P}$, $\mathcal{Q}$, $\mathcal{R}$, and $\mathcal{S}$.
    The insets show the best-fitting power-law parameters with $1\sigma$ error contour.
    The error bars on the markers are the square root of the diagonals of the covariance matrix obtained from the jackknife resampling method.}
    \label{fig:result_wp_selection}
\end{figure*}

In this section, we study the dependence of 2pCF on WISE band selections applied on the $r$-band-limited GAMA data.
As explained in Sect.~\ref{sec:data_w1w4selection}, we applied the W2, W3, and W4 selections in the $r$+W1 selected samples $\mathcal{P}$, $\mathcal{Q}$, $\mathcal{R}$, and $\mathcal{S}$ and computed the 2pCF in the resulting 16 subsamples.
The 2pCF measurements and best-fitting power-law parameters with $1\sigma$ error contours in these 16 subsamples are shown in Fig.~\ref{fig:result_wp_selection}.
Each panel of the figure represents a different redshift range, and different curves in each panel represent different selections as labelled. 
The best-fit parameter values are tabulated in Table~\ref{table:wp_params_selectiondep}.
We find that the $r$+W1+W4 selected galaxies exhibit a weaker clustering than $r$+W1 selected galaxies.

\begin{table}
\caption{Best-fitting power-law parameters for the galaxy samples we used to study the dependence of 2pCF on W1 to W4 selection.}
\begin{center}
\begin{tabular}{l c c c}
\toprule
\toprule
  \multicolumn{1}{l}{Selection} &
  \multicolumn{1}{c}{$r_0 \, (h^{-1} \mathrm{Mpc})$} &
  \multicolumn{1}{c}{$\gamma$} &
  \multicolumn{1}{c}{$\chi^2$} 
  \\ 
\midrule
\multicolumn{4}{c}{Sample $\mathcal{P}$ ($0.07 \leq z < 0.15$)} \\
\midrule
  W1 & $5.65 \pm 0.90$ & $1.67 \pm 0.10$ & 0.01 \\ 
  W2 & $5.67 \pm 0.88$ & $1.67 \pm 0.10$ & 0.01 \\ 
  W3 & $5.24 \pm 0.77$ & $1.63 \pm 0.12$ & 0.01 \\ 
  W4 & $5.04 \pm 0.71$ & $1.62 \pm 0.08$ & 0.01 \\ 
  \midrule
  \multicolumn{4}{c}{Sample $\mathcal{Q}$ ($0.15 \leq z < 0.25$)} \\
\midrule
  W1 & $5.30 \pm 0.37$ & $1.84 \pm 0.05$ & 0.46 \\ 
  W2 & $5.32 \pm 0.39$ & $1.84 \pm 0.05$ & 0.01 \\ 
  W3 & $5.01 \pm 0.36$ & $1.79 \pm 0.05$ & 0.03 \\ 
  W4 & $4.92 \pm 0.42$ & $1.80 \pm 0.06$ & 0.50 \\ 
  \midrule
  \multicolumn{4}{c}{Sample $\mathcal{R}$ ($0.25 \leq z < 0.35$)} \\
\midrule
  W1 & $5.82 \pm 0.35$ & $1.85 \pm 0.06$ & 2.94 \\ 
  W2 & $5.84 \pm 0.32$ & $1.84 \pm 0.05$ & 1.22 \\ 
  W3 & $5.46 \pm 0.32$ & $1.80 \pm 0.05$ & 1.07 \\ 
  W4 & $5.55 \pm 0.32$ & $1.76 \pm 0.06$ & 0.03 \\ 
  \midrule
  \multicolumn{4}{c}{Sample $\mathcal{S}$ ($0.35 \leq z < 0.43$)} \\
\midrule
  W1 & $6.15 \pm 0.40$ & $1.86 \pm 0.06$ & 0.09 \\ 
  W2 & $6.19 \pm 0.38$ & $1.85 \pm 0.04$ & 0.28 \\ 
  W3 & $5.67 \pm 0.34$ & $1.80 \pm 0.06$ & 3.62 \\ 
  W4 & $5.46 \pm 0.36$ & $1.90 \pm 0.05$ & 3.39 \\ 
\bottomrule
\end{tabular}
\end{center}
\label{table:wp_params_selectiondep}
\end{table}

\subsection{Marked correlation functions}\label{sec:result_mcf}

In this section, we present the measurements of MCFs in 16 subsamples described in Table~\ref{table:subsamples_mcf}.
In whole, we computed MCFs using 11 galaxy properties as marks: {\small $L_\text{W1}$}, {\small $L_\text{W2}$}, {\small $L_\text{W3}$}, {\small $L_\text{W4}$}, {\small $M^{\star}_\text{GAMA}$}, {\small $M^{\star}_\text{WISE}$}, {\small $M^{\star}_\text{ProSpect}$}, {\small $\text{SFR}_\text{GAMA}$}, {\small $\text{SFR12}_\text{WISE}$}, {\small $\text{SFR22}_\text{WISE}$}, and {\small $\text{SFR}_\text{ProSpect}$}.

\begin{figure*}[h]
    \centering
    \includegraphics[width=0.9\linewidth]{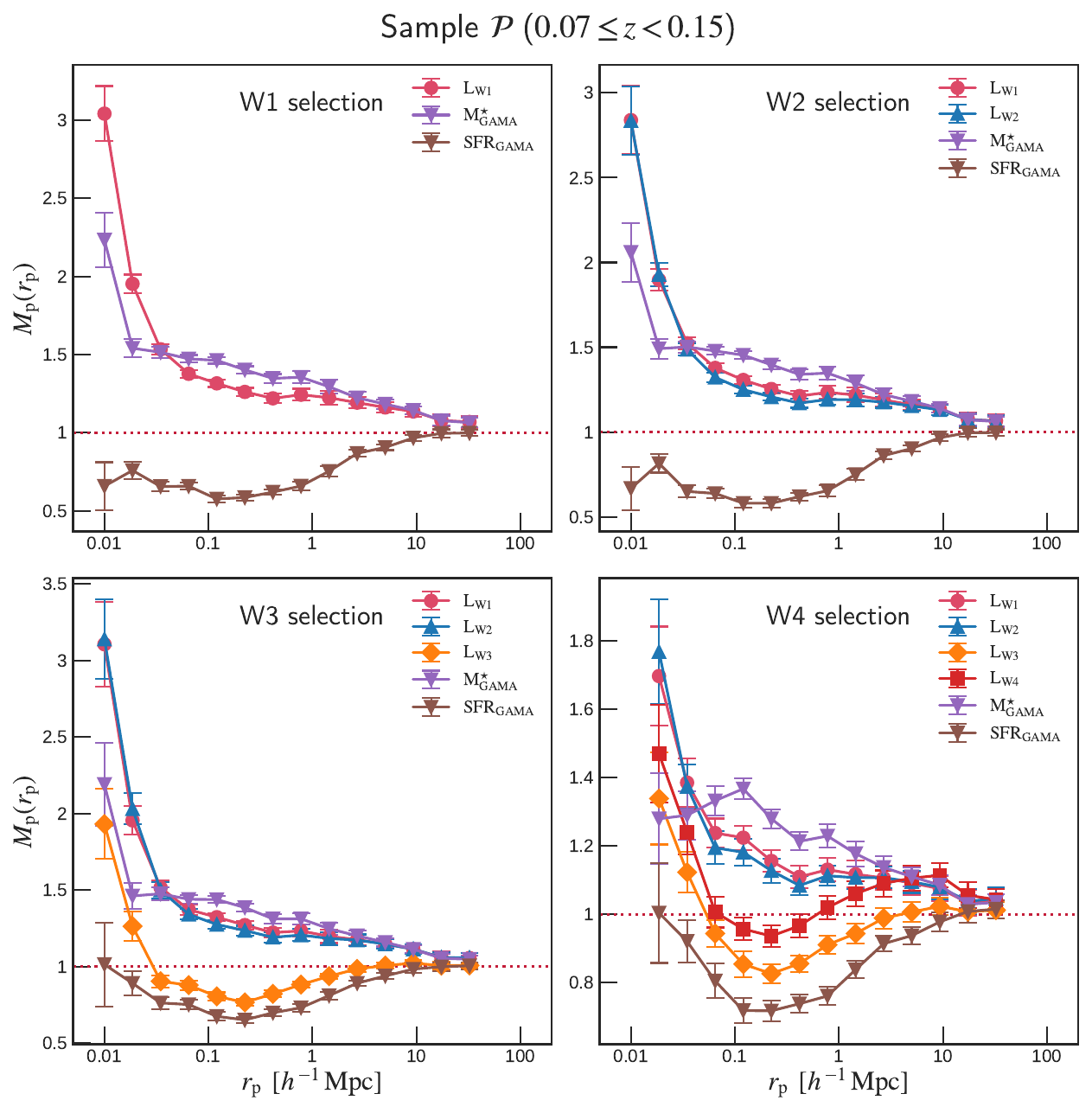}
    \caption{Rank-ordered projected MCFs obtained using different properties (as labelled) in subsamples obtained from the W1 to W4 selection in Sample $\mathcal{P}$.
    The error bars are the combination of jackknife errors and random shuffling errors.}
    \label{fig:result_mcf_p}
\end{figure*}

\begin{figure*}[h]
    \centering
    \includegraphics[width=\linewidth]{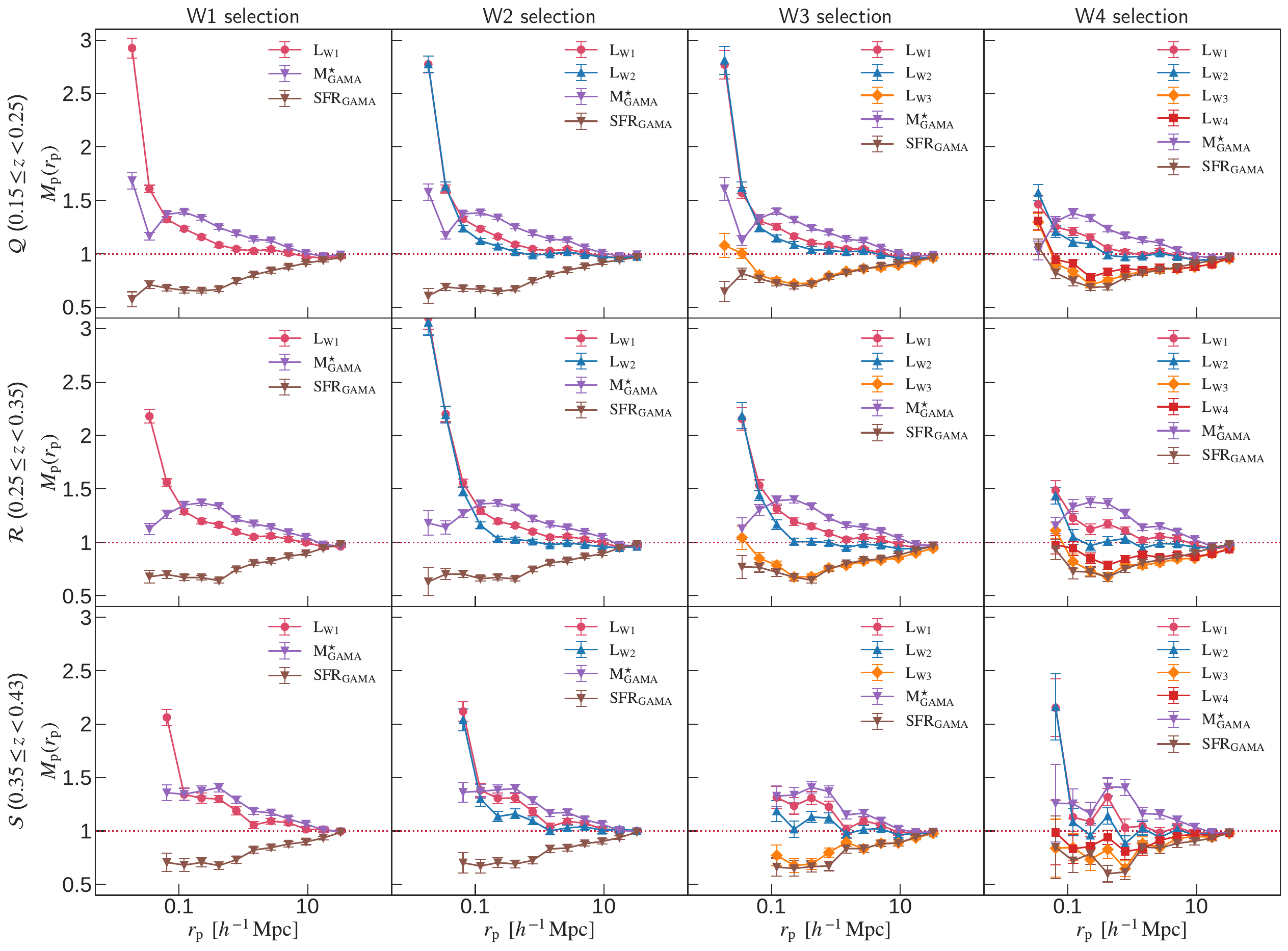}
    \caption{Rank-ordered projected MCFs obtained using different properties (as labelled) in subsamples obtained from the W1 to W4 selection in Sample $\mathcal{Q}$, $\mathcal{R}$, and $\mathcal{S}$.
    Each row represents the samples in the same redshift bin, and each column represents the samples with the same selection.
    The error bars are the combination of jackknife errors and random shuffling errors.}
    \label{fig:result_mcf_qrs}
\end{figure*}

First, we present the {\small $L_\text{W1}$}, {\small $L_\text{W2}$}, {\small $L_\text{W3}$}, {\small $L_\text{W4}$}, {\small $M^{\star}_\text{GAMA}$}, and {\small $\text{SFR}_\text{GAMA}$} MCFs to compare how the mid-IR luminosities follow stellar mass and SFR in their correlation with the environment.
The results in the lower redshift bin (Sample $\mathcal{P}$) are shown in Fig.~\ref{fig:result_mcf_p}, in which each panel represents one of the selections applied on Sample $\mathcal{P}$. 
The results in the rest of the samples are shown in Fig.~\ref{fig:result_mcf_qrs}, where each row shows the measurements in different redshift bins (Samples $\mathcal{Q}$, $\mathcal{R}$, and $\mathcal{S}$), whereas each column deals with the applied selections.
In each selection, we present MCFs with only those properties for which measurements are available in the catalogue (as described in the legends of Fig.~\ref{fig:result_mcf_p}).  
For instance, not for all the galaxies in the $r$+W1 selection are the W4 luminosities measured. 
Therefore, we cannot measure the {\small $L_\text{W4}$} MCF in the $r$+W1 selection.

For a given property (mark), the strength of the deviation of its MCF from unity implies the strength of the correlation (MCF $> 1$) or anti-correlation (MCF $< 1$) between that property and the environment.
In all panels of Fig.~\ref{fig:result_mcf_p} and Fig.~\ref{fig:result_mcf_qrs}, all the MCFs deviate from unity on small scales and approach unity at larger scales.
This shows the environmental dependence of all the properties that we considered for the analysis. 

All the presented MCFs are rank-ordered, that is, the MCFs were measured using the rank of the property in the sample as mark, rather than the value of the property itself (see Sec. \ref{sec:measurement_mcf}).
A simple comparison of the MCFs obtained using different properties can therefore tell us which property is more dependent on environment \citep{skibba2013}.  
Fig.~\ref{fig:result_mcf_p} shows that the stellar mass MCFs ({\small $M^{\star}_\text{GAMA}$}) are the strongest on scales $r_\text{p} > 0.1 \, h^{-1} \, \text{Mpc}$ in all the panels.
The luminosity MCFs {\small $L_\text{W1}$} and {\small $L_\text{W2}$} follow stellar mass MCFs. {\small $L_\text{W3}$} and {\small $L_\text{W4}$} , however, show an opposite pattern and closely follow {\small $\text{SFR}_\text{GAMA}$}.

\begin{figure*}[h]
    \centering
    \includegraphics[width=\linewidth]{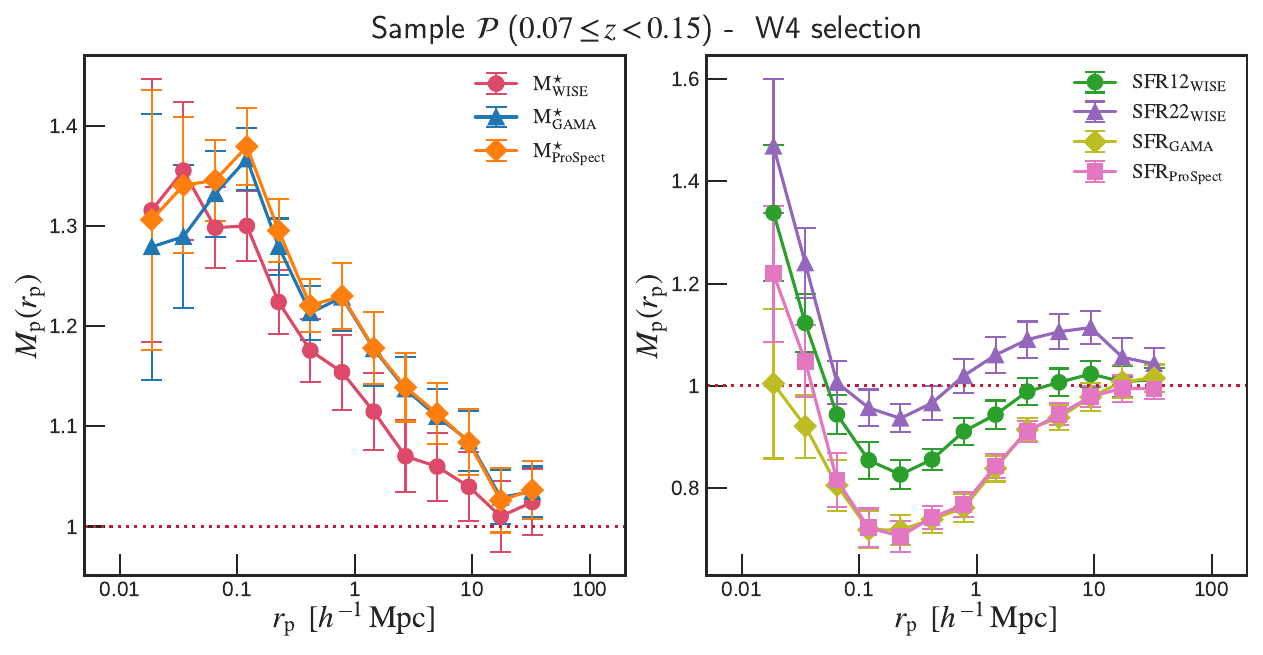}
    \caption{Stellar mass and SFR MCFs in the W4 selection of sample $\mathcal{P}$. \textit{Left:} Stellar mass MCFs obtained using three different estimates of stellar mass as mark.
    \textit{Right:} SFR MCFs with four different estimates of SFR as mark. 
    The error bars are obtained by combining the jackknife errors and random shuffling errors.}
    \label{fig:result_mcf_p_mass-sfr}
\end{figure*}

In Fig.~\ref{fig:result_mcf_p_mass-sfr} we present the stellar mass and SFR MCFs in the W4 selection of sample $\mathcal{P}$. 
The left panel shows the stellar mass MCFs obtained using three different estimates of stellar masses as marks, and the right panel presents the SFR MCFs obtained using four different SFR estimates as marks. 
This observation is discussed in Sect.~\ref{sec:discussion_diffest}.

\section{Discussion}\label{sec:discussion}

\subsection{W1 magnitude dependence of galaxy clustering}\label{sec:discussion_w1_dep}

The two-point correlation function is an efficient tool for studying the clustering of galaxies and its dependence on galaxy properties.
The correlation lengths of different samples can be compared to conclude which sample exhibits stronger clustering: higher amplitude implies stronger clustering (see Sect.~\ref{sec:measurement_powerlaw} for details).
In Sect.~\ref{sec:result_w1dep} we compare the correlation lengths obtained in different samples with varying W1 absolute magnitude limits.

We observe a clear dependence of galaxy clustering on the W1 absolute magnitude in all the redshift bins in the range $0.07 \leq z < 0.43$ (Fig.~\ref{fig:result_w1zr0gam}).
The galaxies that are brighter in the W1 band cluster more strongly than those that are fainter.
This means that in the redshift range of $0.07 \leq z < 0.43$, galaxies that are brighter in the W1 band tend to exist in denser regions of the LSS than their fainter counterparts.
This result agrees with \citet{jarrett2017} as they observed that when the galaxies are divided based on the apparent W1 absolute magnitude, brighter galaxies cluster more strongly than the fainter ones do.
The dependence of clustering on W1 (3.4~$\mu\mathrm{m}$) magnitude can be compared to that on the $K$ (2.2~$\mu\mathrm{m}$) magnitude because the two bands are sensitive to the similarly evolved galaxy population \citep{jarrett2017}.
Many studies showed an enhancement in the clustering amplitude of brighter galaxies in the $K$ band in comparison to the fainter ones \citep{baugh1996, daddi2003, maller2005, sobral2010}.
We note that these wavelengths are dominated by older stellar populations \citep{kauffmann1998}.
The W1 mass-to-light ratio is relatively constant for different stellar populations \citep{wen2013, meidt2014, kettlety2018}.
Galaxies that are brighter in the W1 band are therefore preferably evolved and massive ones and are expected to cluster more due to the stronger clustering behaviour of massive galaxies \citep[e.g.][]{mccracken2015, cochrane2018, durkalec2018}. 

Our observation that galaxies that are more luminous in the W1 band tend to cluster more strongly can be explained in the context of galaxy formation and evolution.
According to the hierarchical model of structure formation, small initial perturbations in the density field of the Universe evolved under gravity and formed the present LSS \citep{press_schechter_1974, mo1996}.
The stronger over-densities resulted in dark matter haloes in which the galaxies were born.
The massive dark matter haloes provided a potential that was strong enough to form the massive galaxies.
This means that the halo mass is tightly correlated with galaxy mass \citep{moster2010}.
Since W1-selected galaxies are preferably massive ones due to their relatively constant mass-to-light ratios \citep{meidt2014}, the clustering of W1-selected galaxies is expected to be driven by the clustering of more massive haloes \citep{conroy2006, wang2007}.

\subsection{Redshift evolution of galaxy clustering}\label{sec:discussion_z_dep}

\begin{figure}[h]
    \centering
    \includegraphics[width=\linewidth]{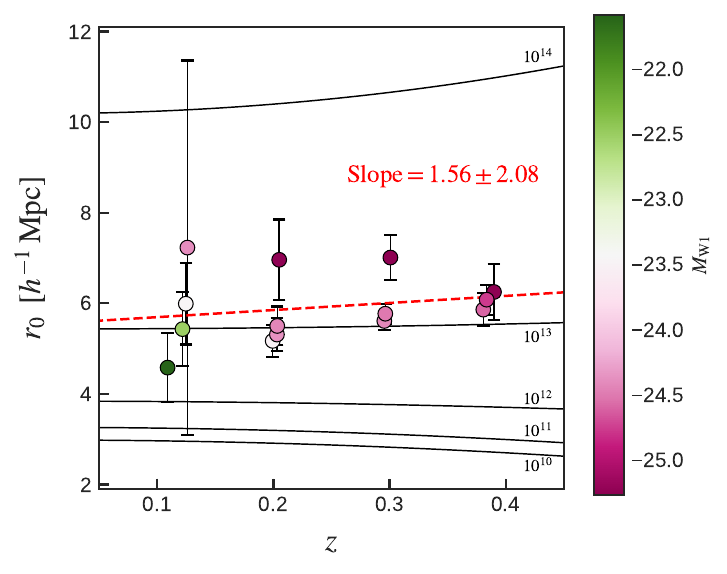}
    \caption{Correlation length $r_0$ as a function of the median redshift of the 14 unique samples described in Sect.~\ref{sec:data_sampleselection_2pcf}.
    The five solid curves show the correlation length of dark haloes with different minimum masses, as labelled.
    The dashed red line represents the best linear fit for the $z - r_0$ relation, whose slope is $1.56 \pm 2.08$.}
    \label{fig:z_r0}
\end{figure}

In Sect.~\ref{sec:result_zdep} we compare the correlation lengths of samples with the same magnitude range, but different redshift ranges.
In Fig.~\ref{fig:z_r0} we show the dependence of the correlation length $r_0$ on the redshift.
The slope of the best-fitting linear function of the $z-r_0$ relation is $1.56 \pm 2.08$.
Therefore, we do not observe a redshift evolution of the clustering of $r$+W1-selected galaxies between $z=0.07$ and $z=0.43$.

We propose two reasons for the lack of a redshift dependence of the clustering of $r$+W1-selected galaxies.
First, the GAMA survey is flux limited with $r < 19.8$.
This imposed flux limit can cause an incompleteness in the samples selected by galaxy properties such as absolute magnitudes and stellar mass.
This effect has been observed in \citet{sureshkumar2021}, where a stellar mass incompleteness imposed by the flux limit impacted the clustering measurements in GAMA.
Since the W1 band traces the stellar mass \citep{jarrett2013}, we expect the same incompleteness in the W1 absolute magnitude, especially in higher redshift bins.
This means that although they have the same W1 absolute magnitude limit, the higher redshift bins select more luminous galaxies than lower redshift bins.
For example, the median $M_\text{W1}$ of the higher redshift sample $\mathcal{N}$4 is $-24.78$, whereas that of the lower redshift sample $\mathcal{N}$1 is $-24.40$ (middle panel of Fig.~\ref{fig:subsamples}).
Therefore, the higher redshift samples exhibit stronger clustering.
This effect counteracts the expected intrinsic evolution of galaxies \citep[e.g.][]{arnouts1999} and might be responsible for the lack of evolution of clustering in our observation.

The second reason might be the lack of significant evolution of host halo mass in the redshift range we considered ($0.07 \leq z < 0.43$).
The redshift evolution of the correlation length can be used to infer information about the host halo mass using the linear galaxy bias model \citep{mo&white2002}.
In Fig.~\ref{fig:z_r0} we show the prediction for the redshift evolution of $r_0$ for different dark matter halo minimum masses (solid black lines). 
These predictions are based on the linear model for the large-scale galaxy bias by \citet{mo&white2002}. 
Based on these predictions, the $r$+W1-selected galaxies in the redshift range $0.07 \leq z < 0.43$ reside in dark matter haloes with an average mass of $\sim10^{13} M_\odot$.
Since the dark matter halo mass does not evolve significantly in this redshift range, we do not expect the clustering strength to show strong evolution.

We also comment on the poorly constrained power-law fit parameters measured for the $\mathcal{N}$1 sample.
The large error contour of this sample is due to the stronger off-diagonal elements of the covariance matrix (Eq.~\ref{eqn:covmat_jackknife}) that are caused by the finite-volume effect.
Our volume-limited samples at different redshift bins span different volumes of the LSS: the lowest redshift slice covers the smallest volume, and the highest redshift sample covers the largest volume.
As a result, the low-redshift samples tend to miss the brighter galaxies.
Since brighter galaxies are expected to be more strongly clustered, this effect distorts the correlation function at the smallest $r_\text{p}$ scales.
In the inset of the right panel of Fig.~\ref{fig:result_zdep}, this effect is visible as a decrease in the slope of the 2pCF of $\mathcal{N}$1 in comparison with other samples (although within the error bars).
This effect was also observed  by \citet[][see Fig.~9]{zehavi2011} in their brightest galaxy sample.

Fig.~\ref{fig:z_r0} also shows that the W1 luminous galaxies tend to occupy massive dark matter haloes.
This qualitatively agrees with the general observation that the most massive galaxies tend to occupy the most massive dark matter haloes.

\subsection{WISE-derived properties as tracers of galaxy clustering}\label{sec:discussion_wisebands}

In Sect.~\ref{sec:result_mcf} we presented the results of MCF measurements in which the galaxies are weighted using the luminosities in W1, W3, W3, W4 bands, stellar masses, and SFRs (see Sect.~\ref{sec:measurement_mcf} for details).
In general we observe two different behaviours among two sets of WISE luminosities. 
Firstly, W1 and W2 luminosity MCFs closely follow the stellar mass MCFs. 
This is not surprising because the W1 and W2 bands are direct indicators of stellar mass as they mainly trace the continuum emission from evolved stars and the major part of baryons in a galaxy are trapped in those evolved stars \citep{jarrett2013}.
This makes W1 and W2 bands nearly ideal tracers of the galaxy stellar mass.
However, the more interesting observation is that the stellar mass W1 and W2 luminosity MCFs are not of same amplitude and there are subtle differences between them. 
This means that even though they are good indicators, W1 and W2 luminosities do not perfectly trace the environmental dependence of stellar mass. 
In other words, a sample selected based on W1 or W2 luminosity does not perfectly exhibit the clustering behaviour of a sample selected based on stellar mass.

Secondly, we observe that the WISE W3 and W4 luminosity MCFs follow the SFR MCFs in general, which is also expected as these bands are indicators of SFR. 
The W3 band at 12~$\mu\mathrm{m}$ is sensitive to PAHs \citep{jarrett2013}, which are abundant in actively star-forming regions of the galaxy \citep{sandstrom2010} and hence a direct estimate of the global SFR \citep{calzetti2007, treyer2010}.
The W4 band (22~$\mu\mathrm{m}$) measures the warm dust continuum, which provides a reliable measure of star formation in the absence of active galactic nucleus (AGN) activity.
However, the connection between W3 and W4 bands and SFR is not direct either: their MCFs do not agree perfectly.
It is to be noted that the SFRs estimated from mid-IR luminosities represent only a part of the total IR SFR.
What is lacking would be the largely obscured star formation in dense molecular cores, which is best traced by the far-IR bands \citep{lacey2008, jarrett2013}.
That is, mid-IR luminosities alone do not provide a complete picture of the star formation. \citet{sureshkumar2021} observed a similar relation between $u$ -band luminosity and SFR. 
These two results lead to the conclusion that monochromatic indicators are not the perfect proxies of the SFR in the context of galaxy clustering.
These indicators may miss parts of the clustering signal that is connected to the total star formation activity in the galaxies.
At this point, it is interesting to check the impact of estimation techniques of stellar mass and SFR on the clustering measurements.
We discuss this in Sect.~\ref{sec:discussion_diffest}.

\subsection{Environmental dependence of different estimates of stellar mass and SFR}\label{sec:discussion_diffest}

In Fig.~\ref{fig:result_mcf_p} and Fig.~\ref{fig:result_mcf_qrs}, we observed that the rank-ordered stellar mass MCF shows the strongest amplitude compared to other properties in the scales $r_\mathrm{p} > 0.1 \, h^{-1} \mathrm{Mpc}$.
This is in agreement with the conclusions of \citet{sureshkumar2021}, in which we compared the stellar mass MCFs to $u, g, r, J,$ and $K$ -band luminosity and SFR MCFs and observed a stronger amplitude for the stellar mass MCF than others.
In this work, we observe that in case of $r$+W1-selected galaxies, stellar mass is the property that more strongly correlates with galaxy clustering than luminosities and SFR.
In other words, in $r$+W1-selected galaxies, more massive galaxies tend to be more strongly clustered.

This agrees with our understanding of  hierarchical structure formation theory, where the galaxy stellar mass is tightly connected to the halo properties \citep{moster2010, wechsler2018} and the halo properties are further correlated with the environment \citep{lemson1999, sheth&tormen2004}.
These two correlations, that is, those between stellar mass and the halo properties and between halo properties and the environment, prompt a tighter correlation between stellar mass and the environment.
In other words, the tight correlation between the stellar mass and halo properties could be the driver behind the correlation between stellar mass and the environment.

When we compare the MCFs computed using stellar masses from different estimates in the W4 selection of Sample $\mathcal{P}$ (left panel of Fig.~\ref{fig:result_mcf_p_mass-sfr}), we observe that the MCFs with GAMA masses and ProSpect masses closely agree with each other.
However, the {\small $M^{\star}_\text{WISE}$} MCFs exhibit a slightly lower amplitude ($1.1\sigma$ on average from {\small $M^{\star}_\text{GAMA}$})  on the scale $r_\mathrm{p} > 0.05 \, h^{-1} \mathrm{Mpc}$.
In panel~(a) of Fig.~\ref{fig:M*-sft-scatter}, we compare these three stellar mass estimates and find a close agreement between {\small $M^{\star}_\text{GAMA}$} and {\small $M^{\star}_\text{ProSpect}$}. 
However, there is a scatter between {\small $M^{\star}_\text{GAMA}$} and {\small $M^{\star}_\text{WISE}$} \citep[also observed by ][]{cluver2014}.
About 94\% of the galaxies shown in panel~(a) of Fig.~\ref{fig:M*-sft-scatter} are star-forming with the colour $\text{W2}-\text{W3} \geq 1.5$ \citep{cluver2014}.
Therefore, the scatter in the stellar mass estimates could be due to the added sensitivity of mid-IR bands to the radiation from dust emission associated with star formation \citep{meidt2012, jarrett2013}.
These star-forming galaxies could dilute the clustering signal and could hence be the reason for the disagreement of {\small $M^{\star}_\text{WISE}$} MCF with other mass estimates.
We therefore conclude that the GAMA and ProSpect stellar masses trace the galaxy environment similarly, but the WISE stellar mass is not as strong as GAMA and ProSpect in tracing the environment.

For the SFR (right panel of Fig.~\ref{fig:result_mcf_p_mass-sfr}), we observe that all the MCFs (except for {\small $\text{SFR22}_\text{WISE}$}) take values lower than unity for a separation scale in the range $0.05 < r_\mathrm{p} < 60 \, h^{-1} \mathrm{Mpc}$.
This implies that the SFR is anti-correlated with the galaxy local density, and there is a small probability of finding pairs of actively star-forming galaxies in the dense regions of the LSS.
This observation is in line with previous findings that passive galaxies tend to cluster more strongly than star-forming galaxies in the low-redshift Universe \citep[e.g.][]{lin2012, wang2013, bethermin2014}.

This can be explained in the context of galaxy evolution.
As galaxies evolve, their hot gas reservoir is removed as a result of different galaxy processes such as ram-pressure stripping \citep{gunn1972}, galaxy harassment \citep{moore1999_harassment}, and tidal disruptions \citep{merritt1983}.
This results in the suppression of star formation activity in the dense environment \citep{lewis2002, gomez2003}.

The anti-correlation between SFR and the environment can also be due to AGN activity.
It is known that massive galaxies tend to host an AGN \citep{wang2008, magliocchetti2020}.
Due to the stronger clustering tendency of massive galaxies, the denser regions of the LSS tend to have more AGN activity that expel gas from its host galaxy, thereby terminating the star formation in the galaxy \citep{cheung2016, argudo-fernandez2016}.

However, we observe an increase in the amplitude of SFR MCFs on scales $r_\mathrm{p} < 0.2 \, h^{-1} \mathrm{Mpc}$.
Enhancement of star formation on these scales is expected due to the tidal interactions \citep{kennicutt1998, li2008, wong2011}.
The rise in SFR MCF on scales $r_\mathrm{p} < 0.2 \, h^{-1} \mathrm{Mpc}$ is most likely evidence of small-scale galaxy interaction. 
\citet{gunawardhana2018} also made a similar observation with SFR MCF in their stellar mass-selected samples from the GAMA survey.

A comparison of the four different SFR estimates is shown in panel~(b) of Fig.~\ref{fig:M*-sft-scatter}.
By comparing the MCFs measured using different estimates of SFR, we see that GAMA SFR and ProSpect SFR are correlated with the environment in a similar way (see Fig.~\ref{fig:result_mcf_p}).
The GAMA SFR is based on the \textsc{MagPhys} model \citep{dacunha2008_gama_sfr} applied to \textsc{lambdar} photometry \citep{wright2016}, whereas the ProSpect SFR is based on the \textsc{ProSpect} SED-fitting code \citep{robotham2020} applied to \textsc{ProFound} photometry \citep{bellstedt2020_kids}.
Both these codes use an energy balance technique to interpret the attenuated stellar emission at UV, optical, and near-IR wavelengths consistently with the dust emission at mid-/far-IR and sub-millimetre wavelengths.
These methods use stellar evolution models by \citet{bruzual2003}, assuming a \citet{chabrier2003} initial mass function and a \citet{charlot&fall2000} dust law.
The ProSpect code additionally implements an evolving metallicity model for individual galaxies.

The MCFs with WISE SFRs as marks ({\small $\text{SFR12}_\text{WISE}$} and {\small $\text{SFR22}_\text{WISE}$}), on the other hand, show significantly different amplitudes than those with {\small $\text{SFR}_\text{GAMA}$} (Fig.~\ref{fig:result_mcf_p}). 
The WISE SFRs are derived using the WISE W3 (12 $\mu\mathrm{m}$) and W4 (22 $\mu\mathrm{m}$) band luminosities calibrated to the total IR luminosity \citep{cluver2017}.
We recall that the W3 and W4 bands of WISE are relatively less sensitive, with the W4 being the least sensitive.
This is reflected in the significant number of S/N$\leq$2 galaxies causing biased GAMA--WISE SFR relations in panels~(c) and (e) of Fig.~\ref{fig:M*-sft-scatter}. 
Therefore, the physical properties of the S/N$\leq$2 galaxies derived out of the W3 and W4 bands are to be used with caution.
Hence, it is currently difficult to directly interpret the MCF signals.
Nevertheless, it is to be noted that despite of the influence of the lack of sensitivity, the {\small $\text{SFR12}_\text{WISE}$} and {\small $\text{SFR22}_\text{WISE}$} MCFs preserve the clustering trend shown by {\small $\text{SFR}_\text{GAMA}$}.

\subsection{Selection effects on galaxy clustering}

In Sect.~\ref{sec:result_2pcf_w1tow4} we observed that the $r$+W1-selected galaxies show a greater value of the correlation length than galaxies with subsequent longer wavelength selections.
W2-, W3-, and W4-selected galaxies in our samples are basically subsets of  the W1-selected sample.
For example, the W4 selection chooses only galaxies that have measurements of W4 luminosity.
The same holds for the W2 and W3 selection.
When a selection in W4 (or W3) is made, we might be selecting the star-forming galaxies that are known to exhibit a weaker clustering in the local Universe \citep{hartley2010, sureshkumar2021}.
This might be the reason why we see a decline in the correlation length when the selection goes from W1 to W4.
However, this observation can also be due to the difference in sensitivities of the WISE bands: the W3 and W4 bands are less sensitive than W1 and W2.
The selection effect that we observe in our samples might therefore be a combined result of the dependence of clustering on the intrinsic WISE-band luminosities and the sensitivity of the survey.

\section{Summary and conclusions}\label{sec:conclusions}

We studied the correlations between WISE properties and galaxy clustering.
We used a set of magnitude-selected galaxy samples from the GAMA-WISE catalogue in the redshift range $0.07 \leq z < 0.43$.
Using MCF, we checked how the WISE bands trace the galaxy clustering.
We compared how stellar masses and SFRs from three different estimates (GAMA, WISE, and ProSpect) trace the environment using MCFs.
Additionally, using 2pCF, we studied the luminosity dependence and redshift evolution of galaxy clustering.

The summary of our main results and conclusions of this work is given below.
\begin{itemize}
    \item We observed a strong dependence of galaxy clustering on the W1 absolute magnitude in the redshift range $0.07 \leq z < 0.43$ in Fig.~\ref{fig:result_w1zr0gam}.
    Galaxies brighter in the W1 band exhibit stronger clustering than their fainter counterparts. 
    At the same time, we did not observe a significant redshift evolution of galaxy clustering in this redshift range (Fig.~\ref{fig:z_r0}).
    \item By comparing the amplitudes of rank-ordered MCFs, we concluded that in $r$+W1-selected samples, stellar mass is more strongly correlated with environment than luminosities and SFR (Fig.~\ref{fig:result_mcf_p} and Fig.~\ref{fig:result_mcf_qrs}).
    \item We showed that the W1 (3.4~$\mu\mathrm{m}$) and W2 (4.6~$\mu\mathrm{m}$) band luminosities can be the best choices among all the WISE bands after stellar mass for tracing the galaxy clustering.
    However, this proxy relation is not perfect, and these bands do not completely catch the clustering signal traced by stellar mass.
    Similarly, the W3 (12~$\mu\mathrm{m}$) and W4 (22~$\mu\mathrm{m}$) bands closely but not entirely follow the trend of the environmental dependence of SFR.
    \item We observed a general agreement between the clustering dependence of galaxy properties estimated using SED fitting techniques such as \textsc{MagPhys} and \textsc{ProSpect}.
    Despite the influence of low S/N sources in the higher-wavelength WISE bands, the corresponding derived properties preserve the general trend of their clustering dependence.
    \item We observed a weak dependence of clustering on the selection we applied on the sample (Fig.~\ref{fig:result_wp_selection}). 
    The W4-selected galaxies exhibit a weaker clustering in the redshift range we considered.
\end{itemize}

Our MCF measurements are the first of their kind made with the WISE-band properties.
In the near future, more IR data will be available through \textit{Euclid} \citep{laureijs2011} and the Nancy Grace Roman Space Telescope \citep{spergel2015_wfirst}.
The future scope of this work includes extending our analysis to these data.
We also intend to study the environmental dependence of galaxy properties in the CosmoDC2 \citep{korytov2019} mock catalogue from the Vera C. Rubin Observatory \citep[][]{lsst_2009}.

\begin{acknowledgements}
We thank the referee Matthieu Béthermin for the useful comments and suggestions.
We thank Sean Lake for providing data points of the correction of redshift evolution of WISE luminosities, Thomas Jarrett and Miguel Figueira for discussion.
U.S. is supported by Jagiellonian University DSC grant 2021-N17/MNS/000045.
U.S. and A.P. are supported by the Polish National Science Centre grant UMO-2018/30/M/ST9/00757.
A.D. is supported by the Polish National Science Centre grant UMO-2015/17/D/ST9/02121.
M.B. is supported by the Polish National Science Center through grants no. 2020/38/E/ST9/00395, 2018/30/E/ST9/00698, 2018/31/G/ST9/03388 and 2020/39/B/ST9/03494.
M.E.C. is a recipient of an Australian Research Council Future Fellowship (project No. FT170100273) funded by the Australian Government.
This work is supported by  Polish Ministry of Science and Higher Education grant DIR/WK/2018/12.
GAMA is a joint European-Australasian project based around a spectroscopic campaign using the Anglo-Australian Telescope. The GAMA input catalogue is based on data taken from the Sloan Digital Sky Survey and the UKIRT Infrared Deep Sky Survey. Complementary imaging of the GAMA regions is being obtained by a number of independent survey programmes including GALEX MIS, VST KiDS, VISTA VIKING, WISE, Herschel-ATLAS, GMRT, and ASKAP providing UV to radio coverage. GAMA is funded by the STFC (UK), the ARC (Australia), the AAO, and the participating institutions. The GAMA website is \url{http://www.gama-survey.org/}.
Based on observations made with ESO Telescopes at the La Silla Paranal Observatory under programme ID 179.A-2004.
During this research, we made use of Tool for OPerations on Catalogues And Tables \citep[TOPCAT;][]{taylor2005_topcat} and NASA’s Astrophysics Data System Bibliographic Services.
This research was supported in part by PLGrid Infrastructure and OAUJ cluster computing facility.
\end{acknowledgements}

\bibliographystyle{aa} 
\bibliography{references} 

\begin{appendix}

\section{Luminosity evolution of the W1 band}\label{app:lum_evol}

Galaxies evolve in their properties across the redshifts.
The redshift range we covered in this work demands a correction for the galaxy luminosities to account for their redshift evolution.
This makes it more convenient to compare the clustering behaviour of samples at different redshift ranges.

\begin{figure}[h]
    \centering
    \includegraphics[width=\linewidth]{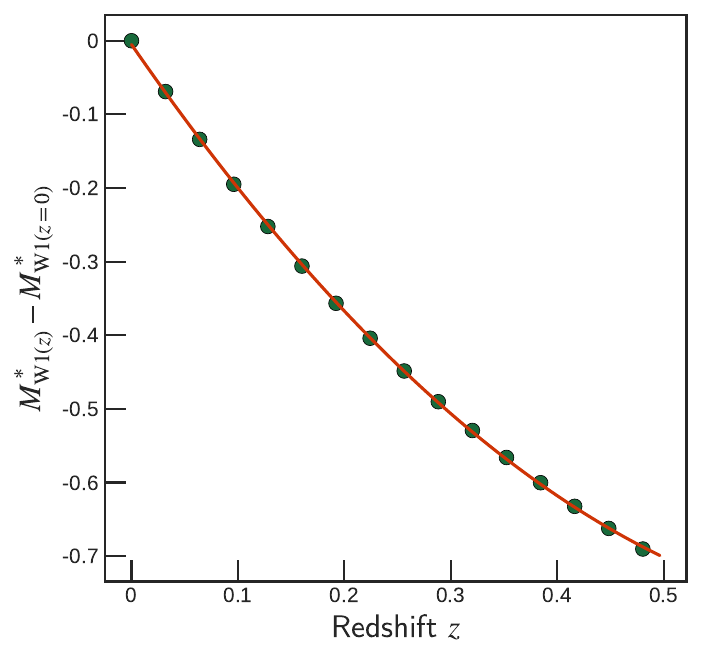}
    \caption{Green circles are the selected points extracted from the dotted black curve in Fig. 7(a) of \citet{lake2018}. 
    The red curve is the polynomial fit given by Eq.~\ref{eqn:fit}.}
    \label{fig:poly_fit}
\end{figure}

For the same purpose, we considered the evolution of the characteristic absolute magnitude in the W1 band ($M^{*}_{\text{W1}}$) as modelled in \citet{lake2018}.
Using the data points taken from the dotted black curve of the Fig. 7(a) of \citet{lake2018}, we fit a second-order polynomial function to $M^{*}_{\text{W1} (z)} - M^*_{\text{W1} (z=0)}$ given by
\begin{equation}\label{eqn:fit}
    M^*_{\text{W1} (z)} - M^*_{\text{W1} (z=0)} = 1.39 z^2 - 2.08 z
.\end{equation}
The fit is shown in Fig.~\ref{fig:poly_fit}.
Using the function given in Eq.~\ref{eqn:fit}, we corrected the W1 absolute magnitudes of all the galaxies in our sample to $z=0$.
This means that for each galaxy, we took $M_\text{W1}  = M'_\text{W1} - (M^*_{\text{W1} (z)} - M^*_{\text{W1} (z=0)}),$ where $M'_\text{W1}$ is the uncorrected W1 absolute magnitude.
\end{appendix}

\end{document}